\documentclass[iop]{emulateapj}
\usepackage{apjfonts}
\usepackage{natbib}
\usepackage{amsfonts}
\usepackage{amssymb,amsmath}
\usepackage[usenames,dvipsnames]{color}
\usepackage{verbatim}
\newcommand\msun{$M_\odot$}

\newcommand{\mathbi}[1]{\textbf{\em #1}}

\newcommand{\epssclone}{\epsscale{1.0}}

\shorttitle{}
\shortauthors{}

\begin{document}

\title{THE DOMINANCE OF NEUTRINO-DRIVEN CONVECTION IN CORE-COLLAPSE SUPERNOVAE}

\author{Jeremiah W. Murphy\altaffilmark{1}}
\author{Joshua C. Dolence\altaffilmark{1}}
\author{Adam Burrows\altaffilmark{1}}

\altaffiltext{1}{Princeton University, Princeton, NJ, USA; jmurphy@astro.princeton.edu,
  jdolence@astro.princeton.edu, burrows@astro.princeton.edu}

\begin{abstract}
Multi-dimensional instabilities have become an important ingredient in
core-collapse supernova (CCSN) theory.  Therefore, it is necessary to
understand the driving mechanism of the dominant instability.
We compare our parameterized three-dimensional CCSN simulations with
other buoyancy-driven simulations and propose scaling relations for
neutrino-driven convection.  Through these comparisons, we infer that
buoyancy-driven convection dominates post-shock turbulence in our simulations.
In support of this inference, we present four major results. First, the convective fluxes and kinetic energies in the
neutrino-heated region are consistent with expectations of
buoyancy-driven convection.  Second, the convective flux is
positive where buoyancy actively drives convection, and the radial and
tangential components of the kinetic energy are in rough equipartition
(i.e. $K_r \sim K_{\theta} + K_{\phi}$).  Both results are natural
consequences of buoyancy-driven convection, and are commonly observed
in simulations of convection. Third,
buoyant driving is balanced by turbulent dissipation. 
Fourth, the convective luminosity and turbulent dissipation
scale with the driving neutrino power.
In all, these four results suggest that in neutrino-driven explosions, the
multi-dimensional motions are consistent with neutrino-driven
convection.

\end{abstract}


\section{Introduction}
\label{introduction}

The explosive death of massive stars, in particular core-collapse
supernovae (CCSNe), are some of the most energetic explosions in the
Universe, and, as such, are fundamental to a wide range of other
astrophysical phenomena.  To highlight a few important examples,
CCSNe are a major site for nucleosynthesis, mark the birth of neutron
stars and black holes, and are major contributors to galactic dynamics
and star formation.  Despite their importance, understanding the mechanism remains an important unsolved problem.  Whatever the
mechanism, it has long been suggested that neutrinos and multi-dimensional
instabilities play major, if not central, roles
\citep{epstein79,bethe85,burrows87,wilson88,bethe90,herant92,benz94,herant94,burrows95,janka96,blondin03,marek09b,murphy08b,nordhaus10}.
In this paper, we use analytic scalings and numerical simulations to
assess whether the dominant multi-dimensional instability is
consistent with neutrino-driven buoyant convection.

Multi-dimensional simulations have long suggested that
aspherical, nonlinear instabilities play important roles in aiding the
delayed-neutrino mechanism toward successful explosions.  
Otherwise, except for the least massive stars
\citep{kitaura06,burrows07d}, the spherical delayed-neutrino
mechanism fails to produce explosions \citep{liebendorfer01a,liebendorfer01b,rampp02,buras03,thompson03,liebendorfer05b,sumiyoshi05,lentz12}.  Even though
the importance of multi-dimensionality is clear, which instability
dominates the aspherical motions has been less clear.  Initially,
neutrino-driven convection was identified as the most relevant
multidimensional instability
\citep{burrows87,wilson88,bethe90,benz94,burrows95,janka95}, but then
idealized two-dimensional (2D) simulations discovered
a new instability, the standing accretion shock instability (SASI) \citep{blondin03}.  Both must exist at some level, but it has never
been made clear which dominates in Nature.

Investigating the importance of neutrino-driven convection in CCSN
theory has a long history.  In the earliest investigations, it was
suggested that convection expands the shock radius, making the
gain region larger and increasing net neutrino heating \citep{benz94,burrows95,janka95,janka96}. However, none
of these investigations verified that the turbulent ram pressure is
actually sufficient to expand the shock radius, nor did they verify
that neutrino-driven convection drives turbulence.  More recently, \citet{murphy08b} considered the global conditions for
explosion in parameterized 2D simulations and suggested that turbulence reduces the critical neutrino
luminosity for successful explosions (see also \citet{yamasaki06} and
\citet{murphy11b} for theoretical discussions).  Using parameterized
three-dimensional (3D)
simulations, \citet{nordhaus10} found similar results, but
\citet{hanke12} suggest that the SASI, and not convection, might be
more important in aiding successful explosions.

The SASI is an instability of the standing accretion shock that was
first discovered in idealized simulations which purposely neglected
neutrinos to suppress
convective instabilities \citep{blondin03}.  These idealized
2D simulations exhibited strong up-and-down sloshing motions of the
shock, leading to an immediate connection to the sloshing shock
motions observed in more realistic 2D simulations.  Consequently, many
subsequent studies focused on the mechanism responsible for the SASI
or postulated that the SASI could lead to successful explosions \citep{blondin03,blondin06,foglizzo06,marek09b,scheck08,foglizzo09,sato09,fernandez10,hanke12}.  Linear theory suggests that an advective-acoustic cycle is the mechanism for the SASI \citep{guilet12,foglizzo12}.
These analyses show that under certain conditions an
advective-acoustic instability in addition to the buoyant instability
may operate in the core-collapse context.
However, to more easily study the SASI, the earliest analyses used idealized
simulations in which buoyancy-driven instabilities were suppressed
\citep{blondin03,blondin06,sato09,foglizzo09}.  

More recently, many more investigations of the SASI include neutrino
heating \citep{ohnishi06, iwakami08, takiwaki12, kuroda12, bruenn13},
but only a few specifically focus on the role of the SASI versus
buoyancy-driven instabilities
\citep{foglizzo06,scheck08,fernandez09,burrows12,mueller12,dolence13,hanke13}.  Using a toy model and linear theory,
\citet{foglizzo06} considered the linear growth of convective
instabilities and found that advection can sweep small-perturbation modes out of
the convectively unstable region before they have time to grow to
nonlinear amplitudes.  Hence, they conclude that a negative entropy
gradient is not enough to drive convective instability; one must also
consider the ratio of the advection time to the local buoyancy
timescale ($\chi$).  For $\chi > 3$, the linear convective instability
succeeds, but for $\chi < 3$, the SASI dominates.  However,
\citet{foglizzo06} cautioned that this analysis is best suited for
linear growth of small perturbations, and if the seed perturbations are
sufficiently large, convection may ensue even if $\chi < 3$.
\citet{scheck08} investigated whether this condition is relevant
in more realistic simulations and found that with small initial
perturbations, the SASI initially appeared to dominate when $\chi < 3$.
However, after $\sim$100 ms, large SASI
perturbations appeared to trigger convection.  With larger, but still
modest initial perturbations $\mathcal{O}(10^{-2}v_r)$, where $v_r$
  is the radial velocity, convection
appeared to dominate at all times.  Given that large convective perturbations in the progenitor
\citep{bazan98,meakin07a} will provide large perturbative
seeds, the latter scenario is more likely.  Based upon linear
analysis, \citet{foglizzo06} concluded that ``advective stabilization
weakens the influence of convection on the largest modes,'' but we
suggest that the multi-dimensional simulations indicate otherwise.

For the first time, \citet{mueller12} and
\citet{hanke13} report strong SASI oscillations and weakened
convection in association with neutrino-driven explosions.  However, this result
seems to be rare even in their set of calculations.  The exceptionally
high accretion rate of one particular progenitor model seems to be
responsible in suppressing convection.  In fact, when \citet{hanke13}
included larger seed perturbations, convection seemed to dominate even
in the model that is seemingly more susceptible to the SASI.
Furthermore, \citet{ott13} performed a 3D simulation of the same progenitor and
found that the turbulent motions seemed to be most consistent with
buoyant-driven convection.  Unfortunately, these results are derived
using varying degrees of approximations to the full problem.  Both used
the same progenitor, general relativity (an approximation in the case of \citet{hanke13}), and a finite temperature EOS, but \citet{mueller12}
and \citet{hanke13} use the ray-by-ray approximation to
neutrino transport and a spherical grid, and \citet{ott13} use a neutrino
leakage scheme and a Cartesian grid.  Considering all of the
approximations and parameters, it seems that for 3D exploding models,
it takes unusually conditioned models to result in the SASI.

\begin{figure*}[t]
\epssclone
\plottwo{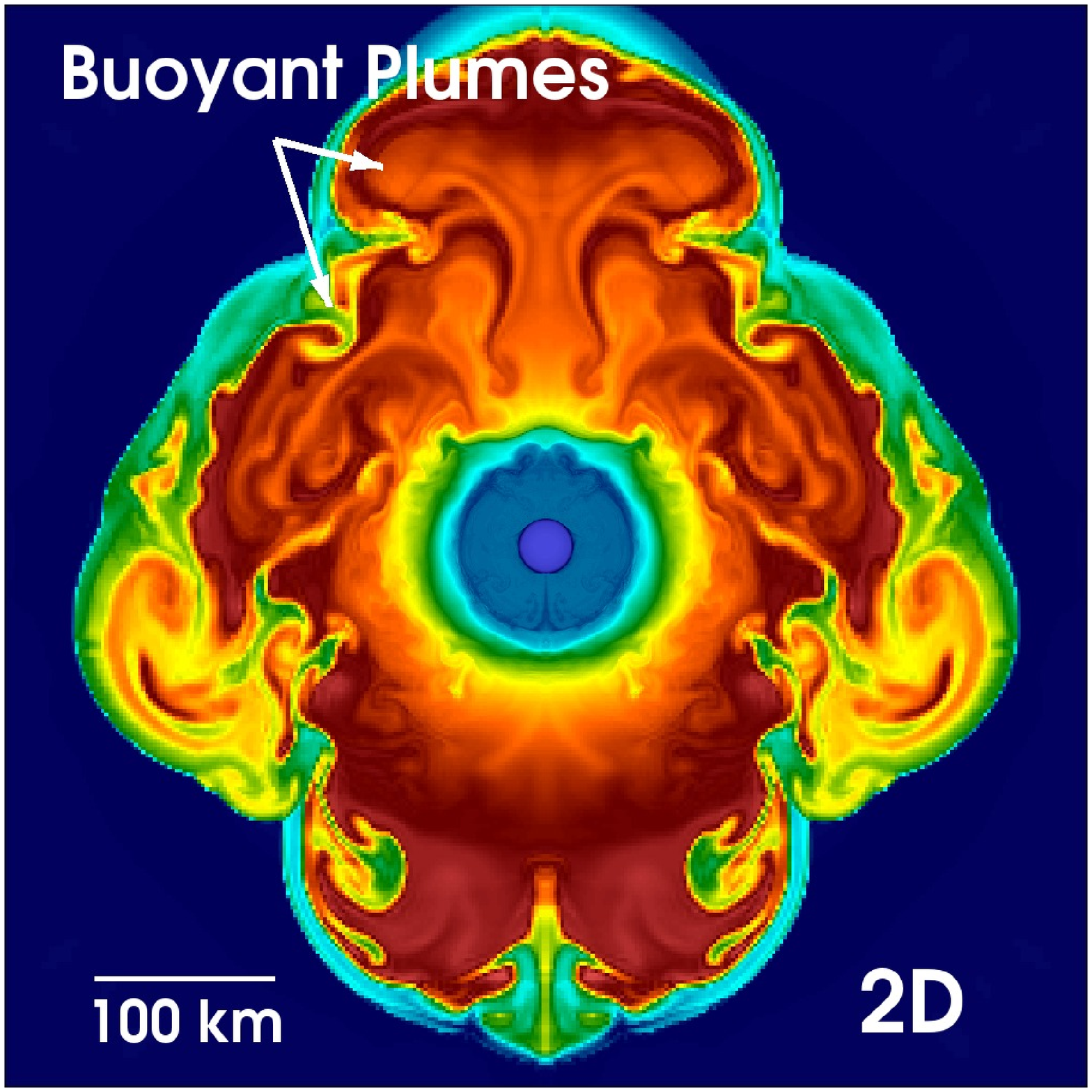}{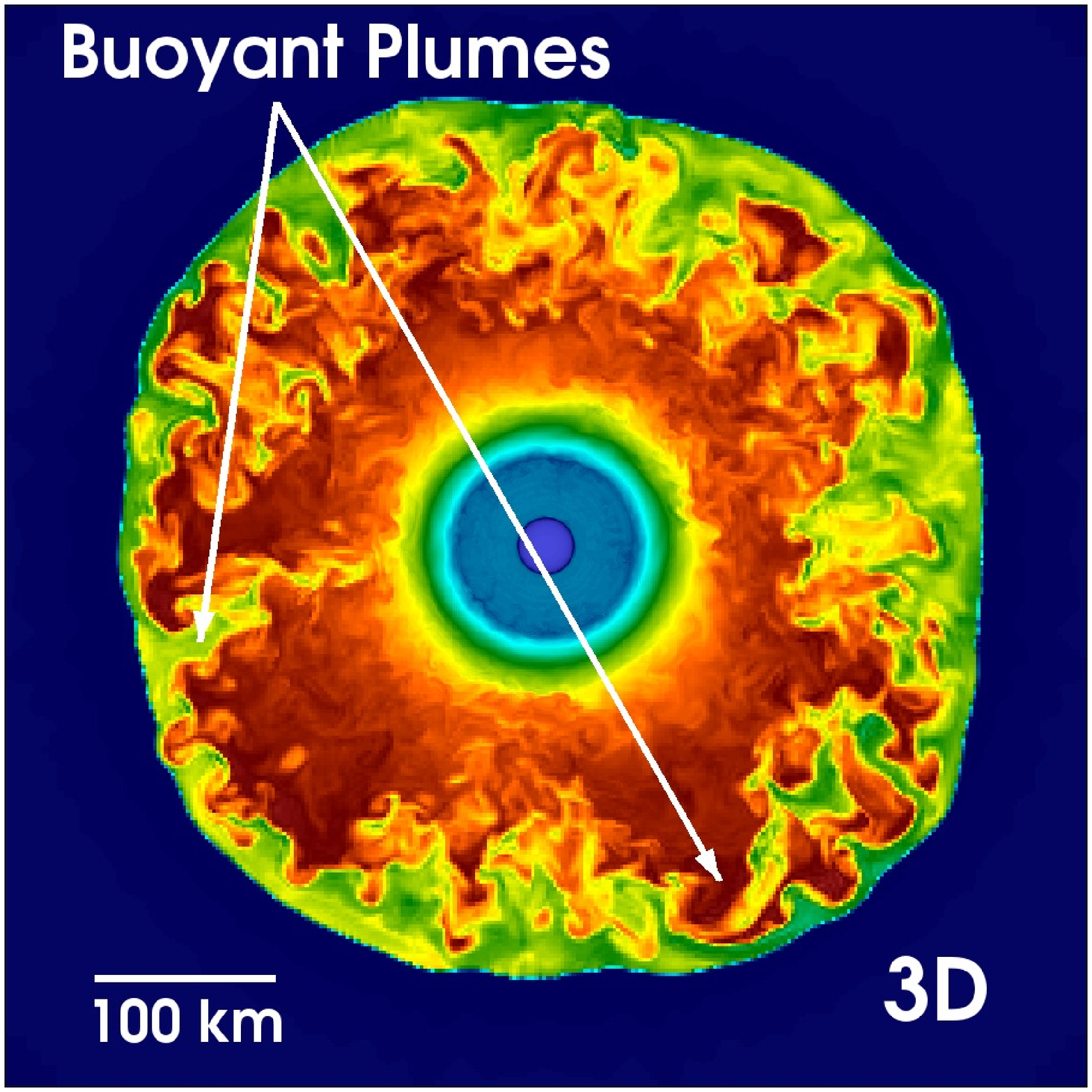}
\caption{Entropy color maps of 2D (left) and 3D (right) CCSN simulations.
  Cooler colors represent lower entropies and warmer colors represent
  higher entropies.  These stills represent the flow at 250 ms after
  bounce for $L_{\nu} = 2.1 \times 10^{52}$ erg/s.  The 2D
  simulation has a higher proportion of coherent structures, which
  turbulence theory predicts \citep{boffetta12}.  Despite the
  differences between 2D and 3D, both show positively (high entropy)
  and negatively (low entropy) buoyant plumes, a strong indication of
  neutrino-driven convection.  \label{plumes}}
\epsscale{1.0}
\end{figure*}

In most simulations, there are hints that buoyancy-driven convection
dominates nonlinear motions.  2D and 3D simulations that
include neutrinos show prominent,
positively-buoyant, high-entropy plumes and negatively-buoyant,
low-entropy plumes at late times (see Figure~\ref{plumes}).  Even in
2D simulations that exhibit large sloshing motions of the shock,
outward excursions of the shock are accompanied by rising, high-entropy
plumes.  Most recently \citet{burrows12} and \citet{dolence13} have analyzed the multi-dimensional
shock and turbulent motions in 2D and 3D simulations and have found that the
sloshing motions frequently identified with the SASI are suppressed in 3D
compared to 2D, and the character of the oscillations is sensitive to
the driving neutrino luminosity.  These results
are consistent with neutrino-driven convection as the source for the
aspherical shock motions, in that the correlation with neutrino luminosity
is an obvious indicator of neutrino-driven convection.  The reduction in the
large-scale sloshing modes in going from 2D to 3D is consistent with known
differences in turbulence between 2D and 3D \citep{boffetta12}.  In 3D,
turbulence cascades to smaller scales only via a constant energy
cascade.  2D turbulence exhibits a double cascade:
an enstrophy cascade to smaller scales and an energy cascade to larger
scales.  This difference naturally leads to more large-scale,
coherent structures in 2D (see Figure~\ref{plumes}).  Might this be the source for the apparent sloshing modes
in realistic 2D simulations?  Albeit circumstantial, these observations 
call into question the assumed dominance of a SASI in
CCSN simulations that include neutrinos.  

Determining which instability dominates, if either,
requires a detailed analysis of the nonlinear motions and comparisons
with theoretical predictions.  Unfortunately, complete nonlinear
theories do not yet exist for either a SASI mechanism or
neutrino-driven convection.  Therefore we can not falsify one theory
or the other; Rather in this paper, we use elements of the incomplete
theory and past numerical experiments to ``derive'' expectations for
the nonlinear turbulence.  The recognition of the SASI is quite recent
and the body of knowledge for the nonlinear SASI is quite limited.  On
the other hand, the body of knowledge related to buoyancy-driven
convection is older and richer.  Therefore, in this paper, we focus on
the latter and leave the former for future work.

To test whether the nonlinear, turbulent flows of our 3D CCSN simulations
are consistent with buoyancy-driven convection, we compare these
simulations with other buoyancy dominated simulations and with expected scalings of neutrino-driven convection.  In
Section~\ref{simulations}, we describe the 2D and 3D
simulations.  Then in Section~\ref{scalings}, we use the
Reynolds-decomposed hydrodynamics equations to formulate scalings
for neutrino-driven convection, and in Section~\ref{scalingsresults}, we
compare these expectations with the properties of 2D and 3D
simulations.  Turbulence, whether it is
neutrino- or SASI-driven, should expand the shock radius.
While this is a trivial prediction, surprisingly, no one has verified
that the shock radius is in fact larger due to this turbulence (as
opposed to increased entropy in the gain region for example).  In
Section~\ref{shockexpansion}, we test whether
the shock radius stalls at larger radii due to turbulence.  Finally, in
Section~\ref{conclusions}, we conclude that the turbulent motions in
these 2D and 3D simulations are consistent with buoyancy-driven
convection.

\section{Simulations}
\label{simulations}

The numerical results of this paper are based upon CCSN simulations
using CASTRO \citep{almgren10} and are similar to the simulations of \citet{burrows12}
and \citet{dolence13}.  CASTRO solves the hydrodynamics equations using a
Godunov-type finite-volume scheme where the boundary fluxes are
calculated using an approximate Riemann solver \citep{almgren10}.
Specifically, it evolves the conservative hydrodynamic equations:
\begin{equation}
\partial_t \rho + \nabla \cdot (\rho {\bf u}) = 0 \, ,
\end{equation}
\begin{equation}
\partial_t (\rho {\bf u}) + \nabla \cdot (\rho {\bf u} {\bf u}) =
-\nabla P + \rho {\bf g} \, ,
\end{equation}
and
\begin{equation}
\partial_t (E) + \nabla \cdot \left [ {\bf u}(E + P) \right ] = \rho
        {\bf u} \cdot {\bf g} + \rho q\, ,
\end{equation}
where $\rho$ is the mass density, ${\bf u}$ is the velocity, ${\bf g}$ is
the local gravitational acceleration, $E$ is $\rho \varepsilon + \rho
u^2/2$, $\varepsilon$ is the specific internal energy, $P$ is the pressure, and
$q$ is the net heating and cooling.  For gravity, we use the
Newtonian monopole approximation, ${\bf g} = -(GM/r^2)\hat{r}$, and for pressure, we use a
relativistic-mean-field equation of state \citep{shen98}.  As initial
conditions for these simulations, we use the 15-\msun\ progenitor model of \citet{woosley95}.\footnote{See \citet{murphy08b} and
\citet{hanke12} for representative accretion rate history curves.}

Following the prescription established in \citet{murphy08b} and
\citet{nordhaus10}, we approximate neutrino heating and cooling with
local prescriptions, i.e.
\begin{equation}
\label{eq:heatcool}
q = \mathcal{H} \, L_{\nu} \left ( \frac{100 \, {\rm km}}{r} \right )^2 \left ( \frac{T_{\nu}}{4 \, {\rm
    MeV}}\right )^2
- \mathcal{C} \left ( \frac{T}{2 \, {\rm MeV}} \right )^6 \,
[{\rm erg/g/s}] \, ,
\end{equation} 
where $L_{\nu}$ is the luminosity of electron- or anti-electron-type
neutrinos in units of $10^{52}$ erg/s, $T_{\nu}$ is the temperature of the neutrinos
(which we set to 4 MeV for all runs), $T$ is the local matter
temperature, and the constants are $\mathcal{H} = 1.544 \times
10^{20}$ and $\mathcal{C} = 1.399 \times 10^{20}$.  For a derivation
of these constants see \citet{janka01}.  In this paper, we consider
neutrino luminosity parameters of $L_{\nu} = 1.5, 1.7, 1.9, 2.1, \, 2.23, \, {\rm
  and}\, 2.3$.

Absorption and emission of
electron- and anti-electron-type neutrinos is most efficient on free
neutrons and protons,respectively.  Therefore, we weight the heating and cooling
terms by the combined mass fractions of protons and neutrons,
i.e. $Y_p + Y_n$.  Equation \ref{eq:heatcool} is an approximation that
is most relevant in the optically-thin regime.  Therefore, to suppress
unphysical heating and cooling at high optical depths, we further
weight Equation~(\ref{eq:heatcool}) by $\exp(-\tau)$, where $\tau = \int{
  \kappa \rho \, dr}$ is an average optical depth of the electron- and
anti-electron-type neutrinos, $\kappa$ is the neutrino opacity, and we
approximate the optical depth with
\begin{equation}
\tau = \frac{3}{4}
\times 10^{-7} \left ( \frac{T_{\nu}}{4\, {\rm MeV}} \right )^2 \int (Y_n + Y_p) \left ( \frac{\rho}{10^{10} \,
  {\rm g/cm^3}} \right ) \, dr
\end{equation}

To simulate the range in length and time scales encountered in core-collapse
simulations, we use CASTRO's adaptive-mesh-refinement (AMR) and adaptive
time-stepping capabilities.  We have developed an AMR strategy to
simulate the full dynamic range of spherical collapse, while
keeping the run-time and memory requirements as low as possible.
Overall, we use six levels of refinement, each a factor of two smaller than
the next largest level.  The
largest domain of the 3D simulations is a cube with 10,000 km on a
side and has a resolution of 32 km at the coarsest level.  To
adequately resolve the proto-neutron star (PNS) structure,
the finest level has a resolution of $\sim$0.5 km out to a radius of 50 km.
In between, we initialize the refinement level ($\ell$) to
maintain a roughly constant angular resolution of $\Delta \theta \sim
0.7^{\circ}$,  i.e. 
\begin{equation}
\label{eq:minlev}
\ell_{\rm min} =
\max \left (
\min \left \{
\left \lfloor
\log_2 \left (
64 \left [ \frac{40 \, {\rm km}}{r} \right ]
\right )
\right \rfloor 
,6 \right \}
,0 \right )
\, ,
\end{equation}
where $\lfloor \cdot \rfloor$ is the floor function.

Throughout the simulation, we maintain Equation~(\ref{eq:minlev}) as the minimum
resolution.   In addition, we set the minimum refinement level to 4
everywhere the entropy is greater than 5 $k_b/{\rm baryon}$ (where
$k_b$ is Boltzmann's constant).
Effectively, this extends level 4 refinement ($\sim$2 km resolution)
to include all regions interior to the stalled shock.  As the shock
expands during explosion, level 4 refinement expands in radius
requiring ever greater memory.  To limit the storage requirements of
the simulation we impose maximum radii for each refinement level via
\begin{equation}
\label{eq:maxlev}
\ell_{\rm max} =
\max \left (
\min \left \{
\left \lfloor
\log_2 \left (
64 \left [ \frac{75 \, {\rm km}}{r} \right ]
\right )
\right \rfloor
,6 \right \}
,0 \right )
\, .
\end{equation}

\section{Scalings for Neutrino-Driven Convection}
\label{scalings}

Comparing to a nonlinear theory for 3D neutrino-driven turbulence is the most robust way to diagnose whether turbulence is driven by buoyancy.   At the moment, a
complete nonlinear theory does not yet exist, but we argue that a
partially complete nonlinear theory is enough to identify the driving
forces of turbulence.  

In particular, we use the Reynolds-decomposed
hydrodynamics equations and formulate scaling relations for
neutrino-driven convection.  In this paper, we are not attempting to
develop a complete theory for neutrino-driven convection.  
Developing a complete theory requires developing a 3D closure model
for the Reynolds-decomposed equations.  Instead, we use the
Reynolds-decomposed equations to diagnose what drives turbulence.  In
short, the decomposed equations clearly delineate the various terms that drive,
dissipate, or redistribute turbulence.  We can use the results of 2D
and 3D simulations to directly calculate the scale of these terms and
determine which dominate.  In this section, we introduce the
Reynolds-decomposed equations and show how we can assess whether or
not buoyancy drives convection in the 2D and 3D simulations of this paper.

The steady-state, spherically-averaged, Reynolds-decomposed
conservation equations for mass, momentum, and entropy are
\begin{equation}
\label{eq:mass}
\nabla \cdot (\rho_0 \mathbi{v} + \left < \rho^{\prime}
\mathbi{v}^{\prime} \right > )
= 0 \, ,
\end{equation}
\begin{equation}
\label{eq:momentum}
\left < \rho \mathbi{u} \right > \cdot \nabla \mathbi{v} = 
-\nabla P_0
+\rho_0 \mathbi{g}
- \nabla \cdot \left < \rho \mathbf{R} \right > \,
\end{equation}
and 
\begin{equation}
\label{eq:entropy}
\left < \rho \mathbi{u} \right > \cdot \nabla s_0 = 
\left < \frac{\rho q}{T} \right >
+ \frac{\rho_0 \epsilon}{T_0}
- \nabla \cdot \left < \mathbi{F}_s \right > \, ,
\end{equation}
where $\left < \cdot \right > $ is an average over solid angle and
approximately one eddy turn-over time, the subscript $0$ denotes the background flow, the prime
denotes the perturbation due to convection, and $\epsilon$ is the
turbulent dissipation.  To avoid cumbersome subscripts later, we do
not use $0$ for the background velocity.  Rather, the background
velocity is ${\bf v}$ and the perturbed velocity is ${\bf
  v}^{\prime}$, i.e. ${\bf u} = {\bf v} + {\bf v}^{\prime}$.  The Reynolds-averaged equations are
similar in form to the usual equations of hydrodynamics, except these
equations have three new terms that are associated with turbulence.
The mass equation, Equation~(\ref{eq:mass}), includes the divergence of the
buoyancy flux, $\left < \rho^{\prime} v^{\prime} \right >$, the
momentum equation includes the divergence of Reynolds stress,\footnote{For practical purposes, we
  calculate $R$ via $\left < \rho u_i v^{\prime}_j\right >/\rho_0$,
  which we find to be nearly identical to $\left <
  v^{\prime}_iv^{\prime}_j \right >$.
  This is because when one expands $\left < \rho u_i v^{\prime}_j
  \right >$ into the individual terms, empirically we find that
  $\rho_0 \left < v^{\prime}_i v^{\prime}_j \right >$ is the dominant
  term.
  Note, we could have easily defined $R$ as $\rho_0 v^{\prime}_i
  v^{\prime}_j$.  However, this definition obscures the behavior of
  the turbulent velocities with a steep density gradient.} $\mathbf{R} =
v^{\prime}_iv^{\prime}_j$, and the
entropy equation includes the transport of entropy by the turbulent
entropy flux, $F_s = \rho_0 \left < v^{\prime} s^{\prime} \right >$.  For low-Mach-number flows, the buoyant flux and entropy flux can be related by
a thermodynamic derivative \citep{murphy11b}, so in the rest of this
paper, we consider only $\mathbf{R}$ and $F_s$.  

The new turbulent terms require additional equations to close
the system of equations.  One should refer to \citet{murphy11b} for the full set, but
here we discuss only the equation for $\mathbf{R}$, or more
specifically, we present the specific kinetic energy ($K$) equation, where
$K$ is related to the trace of the Reynolds stress by $K =
(1/2){\rm Tr}(\mathbf{R})$.  The turbulent kinetic energy
equation
is
\begin{equation}
\label{eq:kinetic}
\begin{array}{llll}
\lefteqn{ \partial \langle \rho K \rangle / \partial t
+ \nabla \cdot \left ( \left < \rho K \right > \mathbi{v} \right ) = } 
\\
& & & 
- \mbox{Tr}\left ( \left < \rho \mathbf{R}\right > \cdot \nabla \mathbi{v}\right )
+ \left < \rho^{\prime} \mathbi{v}^{\prime} \right > \cdot \mathbi{g}
- \nabla \cdot \left < \mathbi{F}_{K} \right >
- \nabla \cdot \langle \mathbi{F}_P \rangle  
\\
& & & 
+ \left < P^{\prime} \nabla \cdot \mathbi{v}^{\prime}  \right >
- \rho_0 \epsilon \, .
\\
\end{array}
\end{equation}
For a spherically symmetric background flow,
  this equation completely describes the evolution of the turbulent
  kinetic energy.  On the left-hand-side, we have the time rate of change
of the turbulent kinetic energy, and the second term represents the
redistribution of the turbulent kinetic energy by the average
background flow.  On the right-hand-side, we have the terms that
govern the evolution: shear driving term, work done by buoyancy, turbulent redistribution
by the turbulent kinetic energy flux, the divergence of the pressure flux ($\mathbi{F}_P =
  P^{\prime} \mathbi{v}^{\prime}$), work done by
turbulent pressure, and turbulent dissipation.

By itself, the turbulent kinetic energy equation,
  Equation~(\ref{eq:kinetic}), does not completely determine the turbulent
flow.  For that, one must solve the full set of equations and develop a
closure model for 3D \`{a} la \citet{murphy11b}.  In the process,
one must validate the full set of equations with 3D simulations,
develop a 3D turbulence model for the third-order moments, and
compare the 2D and 3D turbulence models.  Such a task is beyond the
scope of this paper.
Fortunately, there is no need to develop a full turbulence model to
diagnose whether the postshock turbulence is buoyancy-driven.  Rather,
we argue that the global properties of Equation~(\ref{eq:kinetic}) enable
  such a diagnosis.

Using a few standard assumptions, we now suggest a simple balance law
which the 3D simulations should obey if turbulence is buoyancy driven.
Assuming no shear, steady state, zero turbulent kinetic energy flux at the convective
boundaries, and low Mach-number turbulent flows, we integrate 
Equation~(\ref{eq:kinetic}) over the entire convective volume and find a
balance between global buoyant driving and global turbulent dissipation:
\begin{equation}
\label{eq:globalbalance}
\int \left < \rho^{\prime} \mathbi{v}^{\prime} \right > \cdot
\mathbi{g} \, dV
=
\int \rho_0 \epsilon \, dV \, .
\end{equation}
Under the aforementioned assumptions many of the terms trivially
disappear, resulting in a simple balance between two terms.  Note, the
surface terms associated with the second term on the left-hand-side vanish not because we ignored the
background flow, but because we have assumed that the turbulent
kinetic energy vanishes at the convective boundaries.  If the shock
itself generates turbulence (as might be the case for the SASI), then this assumption may be invalid.  For
the moment, however, we will adopt this assumption and let comparison
with 3D simulations (in Section \ref{scalingsresults}) validate or
invalidate this hypothesis.

In comparing Equation~(\ref{eq:globalbalance}) with 3D
simulations, calculating the integrated buoyant driving term ($W_b$)
is straightforward.  We merely use the simulations to calculate the integral,
$W_b = \int \left < \rho^{\prime} v^{\prime} \right > g \, dV$.
Turbulent dissipation, $E_k = \int \rho_0 \epsilon \, dV$, on the other hand requires a model.
We adopt Kolmogorov's hypothesis, in which the
dissipation rate is set at the largest scales\footnote{Actually, the idea that
  dissipation starts at the largest scales and cascades to smaller
  scales was first proposed by \citet{richardson22}, but \citet{kolmogorov41}
  established the quantitative theory that we reference in this paper.} and is of order
$\epsilon \sim {v^{\prime}}^3/\mathcal{L}$, where $v^{\prime}$ is a typical
turbulent velocity on the largest length scale, $\mathcal{L}$.
Upon initial inspection, it might seem that we have
  merely re-framed our ignorance in the parameter $\mathcal{L}$, and
  that the global balance hypothesis is not a predictive theory.
  However, it is indeed predictive.  Global balance,
  Equation (\ref{eq:globalbalance}), together with Kolmogorov's hypothesis
  specifically predicts that buoyant driving is proportional to the
  third power of the turbulent velocity, i.e.
\begin{equation}
\label{eq:globalbalance2}
\int \left < \rho^{\prime} \mathbi{v}^{\prime} \right > \cdot
\mathbi{g} \, dV
\propto
\int \rho_0  {v^{\prime}}^3\, dV \, .
\end{equation}
This is a nontrivial, falsifiable prediction.  For
  example, in
  shear-driven turbulence, rather than buoyancy, the power in shear would be proportional to
the third power of the turbulent velocity.  In this case, the
proportionality in Equation (\ref{eq:globalbalance2}) would not necessarily
hold.  Even if buoyancy is the dominant driving force, this
proportionality may not hold.  For example, $\mathcal{L}$ may not be constant, in which case simulations
would not exhibit the behavior in Equation (\ref{eq:globalbalance2}).  In
summary, if global buoyant driving is indeed balanced by turbulent
dissipation and the length scale is a constant, then we nontrivially predict the
proportionality in Equation (\ref{eq:globalbalance2}).  In Section
\ref{scalingsresults}, we test this hypothesis with several 3D
simulations spanning a wide range of
driving neutrino luminosities; the simulations validate the hypothesis.

Formally, $\mathcal{L}$ is a free parameter of the model.  However,
simulations of stellar models \citep{arnett09} and core collapse
\citep{murphy11b} indicate that $\mathcal{L}$ takes on the largest
possible value, the radial extent of the region actively driving convection.
For this paper, we find that setting $\mathcal{L}$ to the size of the gain
region satisfies global balance.  In effect, $\mathcal{L}$ is no longer a free
parameter, but a condition imposed by the global structure.
If the post-shock turbulence is driven by buoyancy, then the
3D simulations should be consistent with
Equation~(\ref{eq:globalbalance}).  In Section~\ref{scalingsresults},
  we show that the 3D simulations are not only consistent with the
  proportionality in Equation~(\ref{eq:globalbalance2}), but adopting the
  most natural length scale makes the simulations consistent with the
  global balance hypothesis (Equation~\ref{eq:globalbalance}).

Next, we formulate scaling relations for neutrino-driven
convection.  Because a detailed theory for neutrino-driven convection
does not yet exist, we can not yet derive an analytic theory from
first-principles.  However, when a first principles derivation is out
of reach, it is common practice to use dimensional analysis and
experience to suggest analytic scalings and then test these with either
experiment or numerics.  Kolmogorov's theory for turbulence is a
classic and successful example of using this methodology.  Here, we
use a similar
strategy to derive the scalings for neutrino-driven
convection.  

Our primary hypothesis is that in neutrino-driven
convection, the convective power scales with the driving neutrino
power (i.e. $P_{\rm conv} \propto P_{\nu}$).  What are $P_{\rm conv}$
and $P_{\nu}$?  Well, one might use simple dimensional analysis to
guess at the form of these powers.  Instead, we appeal to the governing
equations to help inform an appropriate expression.  Specifically, we use the integral form of the
entropy equation (Equation~\ref{eq:entropy}).  Our next major hypothesis in
deriving the scalings is that the source terms in
Equation~(\ref{eq:entropy}) are of the same order.  Using these
terms, we suggest analytic scalings for neutrino-driven convection,
and test the resulting predictions with 3D simulations in Section~\ref{scalingsresults}.

Next, we consider the entropy equation to find more appropriate
expressions for $P_{\rm conv}$ and $P_{\nu}$.  The statement that the
source terms in Equation~\ref{eq:entropy} are of the same order is equivalent to
\begin{equation}
\label{eq:anaderive1}
\rho q \sim \rho \epsilon
\sim \frac{T_0}{4 \pi r^2} \frac{\partial L_s}{\partial r} \, ,
\end{equation}
where $L_s \equiv 4 \pi r^2 F_s$.
To express this in terms of $L_{\nu}$, we substitute
the expression for $q$ (Equation~\ref{eq:heatcool}) into this expression,
and assume that cooling is negligibly small in the heating region.
Because the heating term is proportional to $L_{\nu} \kappa / r^2$,
Equation~(\ref{eq:anaderive1}) becomes
\begin{equation}
\label{eq:anaderive2}
L_{\nu} \kappa \rho \sim 4 \pi r^2 \rho \epsilon \sim T_0
\frac{\partial L_s}{\partial r} \, , 
\end{equation}
where $\kappa$ is the opacity to
neutrinos.  An order-of-magnitude integration of this last expression
leads to
\begin{equation}
\label{eq:analytic1}
L_{\nu} \tau  \sim E_k \sim T_0 L_s  \, .
\end{equation}
Therefore, if neutrino-driven convection dominates the turbulent
motions, then we expect the driving neutrino power, $L_{\nu} \tau$,
the maximum of the turbulent luminosity, $T_0 L_s$, and the turbulent
dissipation, $E_k$, to be proportional to one another.  For example,
from these scaling relations we predict that
\begin{equation}
\label{eq:tlsvslnutau}
T_0 L_s = \alpha L_{\nu} \tau \, ,
\end{equation}
where $\alpha$ is some constant of proportionality. 
Furthermore, we can now propose an expression for $P_{\rm conv} =
P_{\nu}$.  Our inspection of the entropy equation suggests that
$P_{\rm conv} = T_0 L_s + E_k$ and that $P_{\nu} = L_{\nu} \tau$.
Note that the natural neutrino-driving power to consider is not just
$L_{\nu}$ but $L_{\nu} \tau$, which takes into account the amount of
neutrino power absorbed in the convective region.  Therefore, our hypothesis
becomes
\begin{equation}
\label{eq:tlsvslnutauek}
T_0 L_s + E_k \sim L_{\nu} \tau \, .
\end{equation}
In Section \ref{scalingsresults}, we find that the 3D simulations are
consistent with the hypotheses in Equations.~(\ref{eq:tlsvslnutau}) and
(\ref{eq:tlsvslnutauek}).

\subsection{Results}
\label{scalingsresults}

\begin{figure}[t]
\epsscale{0.7}
\plotone{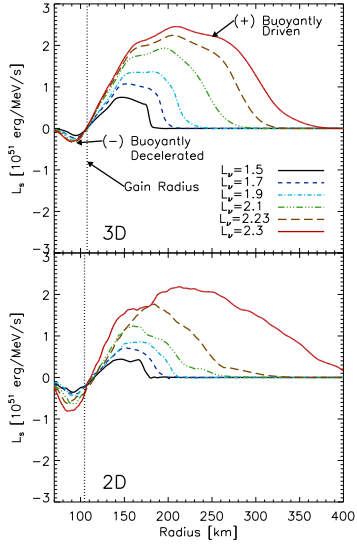}
\caption{Turbulent entropy luminosity , $L_s = 4 \pi r^2 \left < F_s
  \right >$ vs. radius at 250 ms after bounce for 3D (top panel) and
  2D (bottom panel) simulations and six driving
  neutrino luminosities ($L_{\nu} =$ 1.5, 1.7, 1.9, 2.1, 2.23, and 2.3$\times
  10^{52}$ erg/s).  In general, these profiles are consistent with a
  neutrino-driven convection hypothesis.  $L_s$ is positive in the
  gain region where buoyancy actively drives convection, and it is negative where
  stabilizing entropy gradients cause buoyant deceleration.  As is
  expected for neutrino-driven convection, the magnitude of the
  turbulent luminosity monotonically increases with the driving
  neutrino luminosity.\label{lsprofile}}
\epsscale{1.0}
\end{figure}

\begin{figure}[t]
\epsscale{0.6}
\plotone{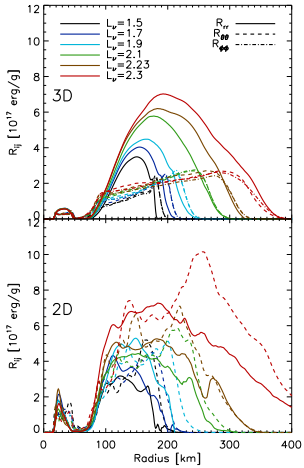}
\caption{Reynolds stress as a function of radius and driving neutrino
  luminosity at 250 ms after bounce for 3D (top panel) and 2D (bottom
  panel) simulations.  
  All three diagonal components are shown:  solid lines correspond to
  $R_{rr}$, dashed lines correspond to $R_{\theta \theta}$, dot-dashed
  lines correspond to
  $R_{\phi \phi}$.  On average, $R_{rr} \sim R{\rm tran}$ globally,
  where $R_{\rm tran}$ is the sum of the transverse components;
  $R_{\rm tran} = R_{\theta \theta} + R_{\phi \phi}$ for 3D, and
  $R_{\rm tran} = R_{\theta \theta}$ for 2D. For 3D, $R_{\phi \phi} \approx R_{\theta
    \theta}$ locally.  This
  equipartition in kinetic energy between the radial and tangential
  components is a commonly observed feature in buoyancy-driven
  convection, and is a consequence of buoyant driving in the radial
  direction, redistribution to the tangential components, and turbulent
  dissipation among all of the components.  As is expected for
  neutrino-driven convection, $R_{rr}$ increases with neutrino
  luminosity.  \label{reyprofiles}}
\epsscale{1.0}
\end{figure}

In this section, we present many ways in which both 2D and 3D CCSN simulations are
consistent with the hypothesis that neutrino-driven convection
dominates the multi-dimensional motions.  Since the 3D simulations
best represent Nature, we highlight the 3D results and only include
the 2D results as a comparison with historical literature.

Figure~\ref{lsprofile} shows the
turbulent entropy luminosity, $L_s = 4 \pi r^2 F_s$, versus radius for six
different driving neutrino luminosities at 250 ms after bounce.  The gross features of these
profiles are consistent with buoyancy-driven convection.
In regions where buoyancy drives convection, higher entropy plumes are driven upward and lower entropy plumes sink.  The
  net effect is that the entropy-and-velocity-perturbation correlation,
  $\left< v^{\prime} s^{\prime} \right>$ (and equivalently $L_s$) is positive in the
  convectively active regions.  As the plumes penetrate into the
  stable layers at the convective boundaries, the entropy
  perturbations flip sign.  For example, sinking plumes, which had a negative
  entropy perturbation have, by definition, positive entropy
  perturbations with respect to the lower bounding layer.  Hence,
  while $L_s$ is positive where buoyancy actively drives convection,
  $L_s$ is negative in the convective overshoot regions.
  \citet{murphy11b} showed that $L_s$ in the gain region of 2D
  CCSN simulations has these characteristics, thereby suggesting that
  convection is buoyantly driven.  Figure
  \ref{lsprofile} shows that
  these same qualities are manifest in similar 3D CCSN simulations;
  once again, suggesting that turbulence in the gain region is
  buoyancy-driven.

Although the $L_s$ profiles of the 2D simulations \citep[see also][]{murphy11b}
and the 3D simulations (this paper) are qualitatively similar, we
note a few interesting and potentially important quantitative differences.
\citet{murphy11b} described the $L_s$ profile as peaking near the gain
radius and sloping down linearly to either side.  Specifically, they
concluded that $L_s$ smoothly approaches zero at the shock.  The
3D profiles in Figure \ref{lsprofile} contradict this conclusion.
Rather than smoothly approaching zero at the shock, $L_s$ is roughly
constant starting at $\sim$30 km above the gain radius all the way up
to the shock; at the shock, $L_s$ discontinuously drops to zero.  We
suspect that the 2D $L_s$ profiles actually show the same behavior as the 3D
$L_s$ profiles and that averaging across the 2D shock causes this
apparent discrepancy.  Note that in the 3D profiles (Figure
\ref{lsprofile}), the width of the shock smooths out the
discontinuity.   In the 2D simulations, the shock exhibits much larger
shock oscillations \citep{burrows12,dolence13}, smoothing the
discontinuity out over a much larger range of radii.  Apparently,
the major differences in the 2D and 3D shock morphology led to an
erroneous description of the $L_s$ profile in \citet{murphy11b}.  This
observation deserves further confirmation (with other codes),
scrutiny, and discussion, but we leave that for future work as this
paper is primarily concerned with whether postshock turbulence is
buoyancy driven and not the differences between 2D and 3D turbulence.
As a final remark on Figure \ref{lsprofile}, the magnitude of $L_s$
monotonically increases with the driving neutrino luminosity.  This is
yet another observation that is expected for neutrino-driven convection.

\begin{figure*}[t]
\epssclone
\plotone{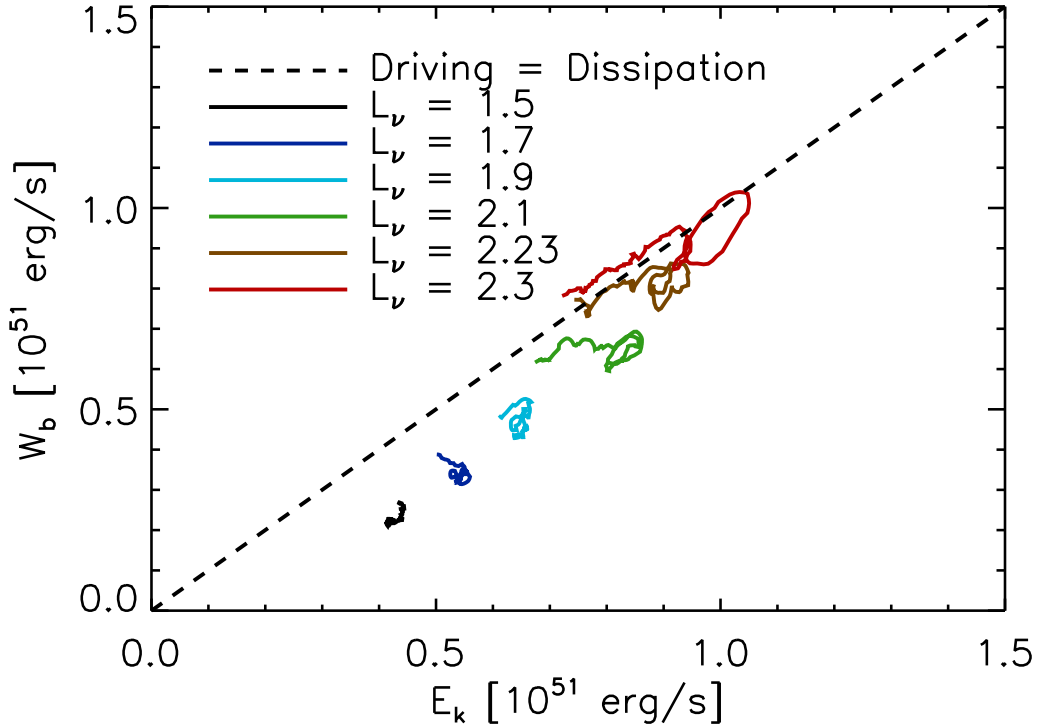}
\caption{Buoyant driving, $W_b$, vs. Turbulent
  dissipation, $E_k$ for 3D simulations.  The fact that buoyant driving is proportional
  to turbulent dissipation and that they are of the same order
  is a strong indicator that the neutrino-driven convection dominates
  the aspherical, nonlinear flow.
  \label{wbvsek}}
\epsscale{1.0}
\end{figure*}

Similarly, neutrino-driven convection explains the Reynolds
stresses.  In Figure \ref{reyprofiles}, we plot the radial ($R_{rr}$,
solid line)
and tangential components ($R_{\theta \theta}$ and $R_{\phi \phi}$,
dashed and dot-dashed lines) of
the Reynolds stress vs. radius at 250 ms after bounce for both the 2D
and 3D simulations.  Two
characteristics of the profiles are consistent with buoyancy-driven convection.  First, like
$L_s$, the strength of the turbulent stresses (mostly
$R_{rr}$) increases monotonically with neutrino luminosity.  Secondly, the radial component of the
turbulent stress is approximately equal to the combined tangential
components; i.e. $R_{rr} \sim R_{\phi \phi} + R_{\theta \theta}$ for
3D and $R_{rr} \sim R_{\theta \theta}$ for 2D.
This result is consistently seen in other numerical experiments where
turbulence is unambiguously driven by buoyancy (see
\citet{arnett09}, and references therein).  In analytic derivations
\citep{arnett09,garaud10,murphy11b}, this approximate equipartition arises
because buoyancy acts first on the radial component, and then the
turbulence is dissipated after energy is redistributed among the three
components.

Even though the Reynolds stress profiles for both 2D an 3D are
consistent with being driven by buoyancy, they are quite different in
detail.  For one, in the outer convective region ($r> 60$ km), the 3D profiles show the largest radial component at
larger radii, but the 2D profiles show the largest components near the
base of the convection zone.  The 2D radial component is also largest
at the base for the inner convective region ($15\, \rm{km} < r < 50\, \rm{km}$), but
is flat in the 3D case.  This implies a major qualitative difference
between 2D and 3D.  In 2D, the peak of $R_{rr}$ at the lowest radii
suggests that $R_{rr}$ is dominated by the sinking plumes.
Conversely, the peak of $R_{rr}$ at the highest radii suggest that
$R_{rr}$ is dominated by the rising plumes.
 

A prediction of the turbulent kinetic energy equation
(Equation~\ref{eq:kinetic}) is that globally buoyant driving is balanced by
turbulent dissipation (Equation~\ref{eq:globalbalance}).  This balance was
also discussed in 3D stellar evolution simulations \citep{arnett09}
and in 2D CCSN simulations \citep{murphy11b}; here, we test whether
buoyant driving is linearly proportional to the turbulent
dissipation and if that constant of proportionality is of order unity.
Figure~\ref{wbvsek} confirms that they are indeed proportional and the
constant of proportionality is of order one.  In this plot, we show global buoyant driving, $W_b$, vs.
global turbulent dissipation, $E_k$, for several driving
neutrino luminosities (delineated by color).  For each
  luminosity, we show $W_b$ and $E_k$ for a range of times from 150 to 260
  ms after bounce.  The time range is chosen to be the same for all
  simulations, late enough to ensure a steady-state stalled shock, and early enough
  to avoid explosion in the highest luminosity run.  As a consequence, each luminosity is not represented by a
single point but a tortured line, which shows that $W_b$ is more or
less balanced by $E_k$ during the entire steady-state period.  In summary, Figure~\ref{wbvsek} verifies that $W_b \propto \int \rho R_{rr}^{3/2} \, dV$,
that the constant of proportionality is the largest scale as predicted
by Kolmogorov, and that global buoyant driving roughly balances global
turbulent dissipation.

Though buoyant driving is proportional to $E_k$, it does
  not exactly equal it.  At the highest luminosity, near explosion they
  are nearly equal, but at the lowest luminosity, $W_b$ is roughly
  half of $E_k$.  There are several possible reasons for this lack of
  exact equality.  First, the definition of $E_k$ is suggested by
  dimensional analysis and is not an exact derivation.  For example,
  even though the size of the gain region may be a good estimate for
  $\mathcal{L}$ at the highest luminosities, it may not be at the
  smallest luminosities.  Second, in deriving Equation~(\ref{eq:globalbalance}),
  we assumed for simplicity that the turbulent kinetic energy fluxes
  at the shock are zero.  The fluxes may not be exactly
  zero\footnote{As might be the case for the SASI.}, but
  small for large luminosities and relatively large for small
  luminosities.  In either scenario, the fact that $W_b$ and $E_k$ are
proportional and are of the same order suggests that turbulent
dissipation is mostly balanced by buoyant driving, and whatever is
missing is a slight correction to this basic conclusion at the lowest
luminosities.  This might be an indication of substantial SASI motions
at the lowest luminosities \citep{burrows12,dolence13}.

\begin{figure*}[t]
\epssclone
\plotone{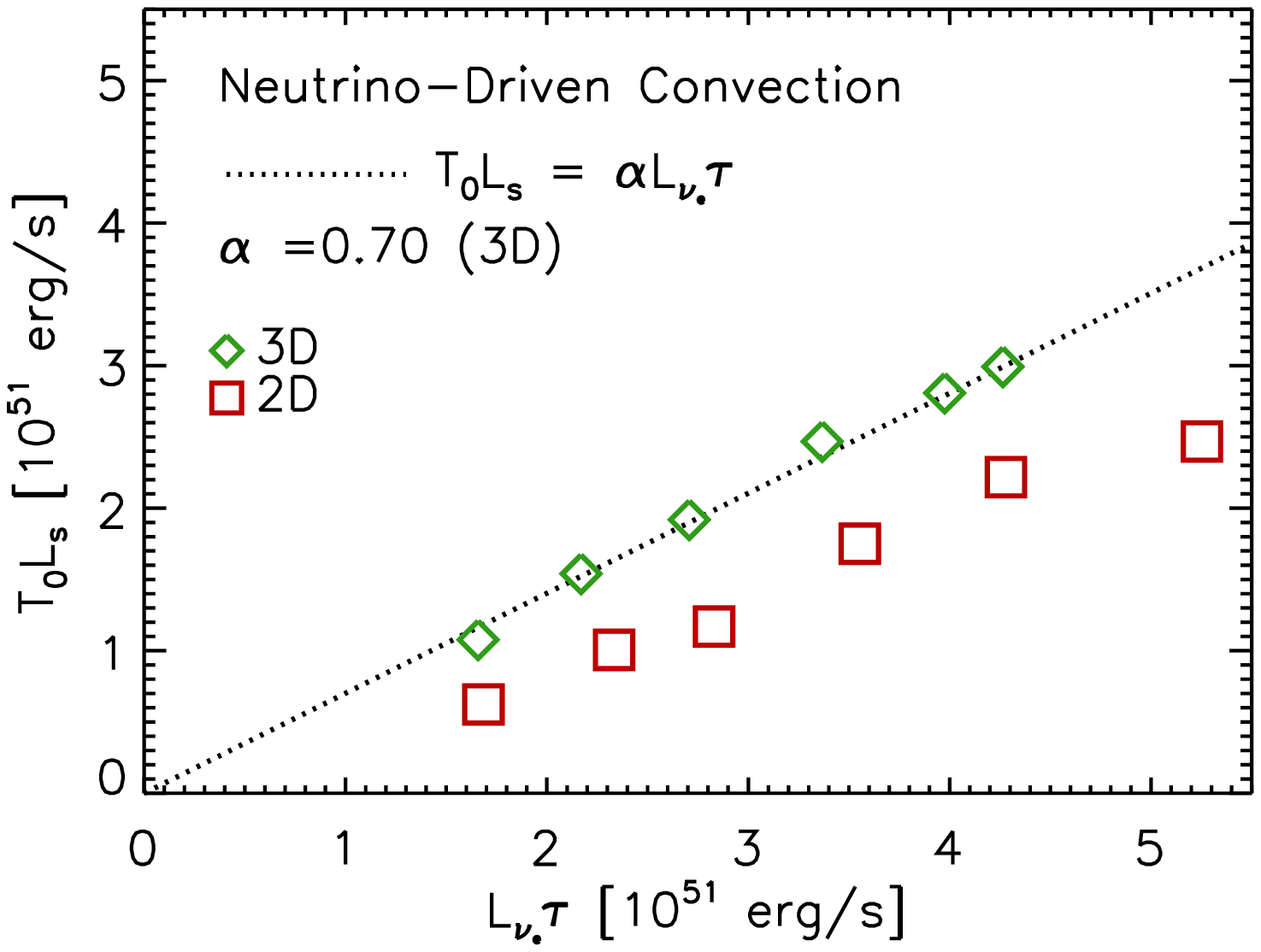}
\caption{Convective luminosity ($T_0L_s$) vs. the driving neutrino
  power, $L_{\nu} \tau$.  The
  symbols show the maximum value of $T_0 L_s$ (restricted to the gain
  region) for six 2D (squares) and 3D (diamonds) simulations, all at 250 ms after bounce.
  For neutrino-driven convection, we analytically expect this
  convective luminosity to be linearly proportional to the driving
  neutrino power.  For comparison, the dashed line shows this linear
  expectation for 3D.  Our analytic calculation does not determine the
  constant of proportionality, so using the 3D simulations, we find that
  $\alpha \sim 0.7$.  The constant for 2D is lower, suggesting less
  efficient driving of convection for 2D.  See Equation~(\ref{eq:tlsvslnutau}) and the associated
  text for the derivation of the analytic scalings with neutrino luminosity.   \label{tlsvslnutau}}
\epsscale{1.0}
\end{figure*}

\begin{figure*}[t]
\epssclone
\plotone{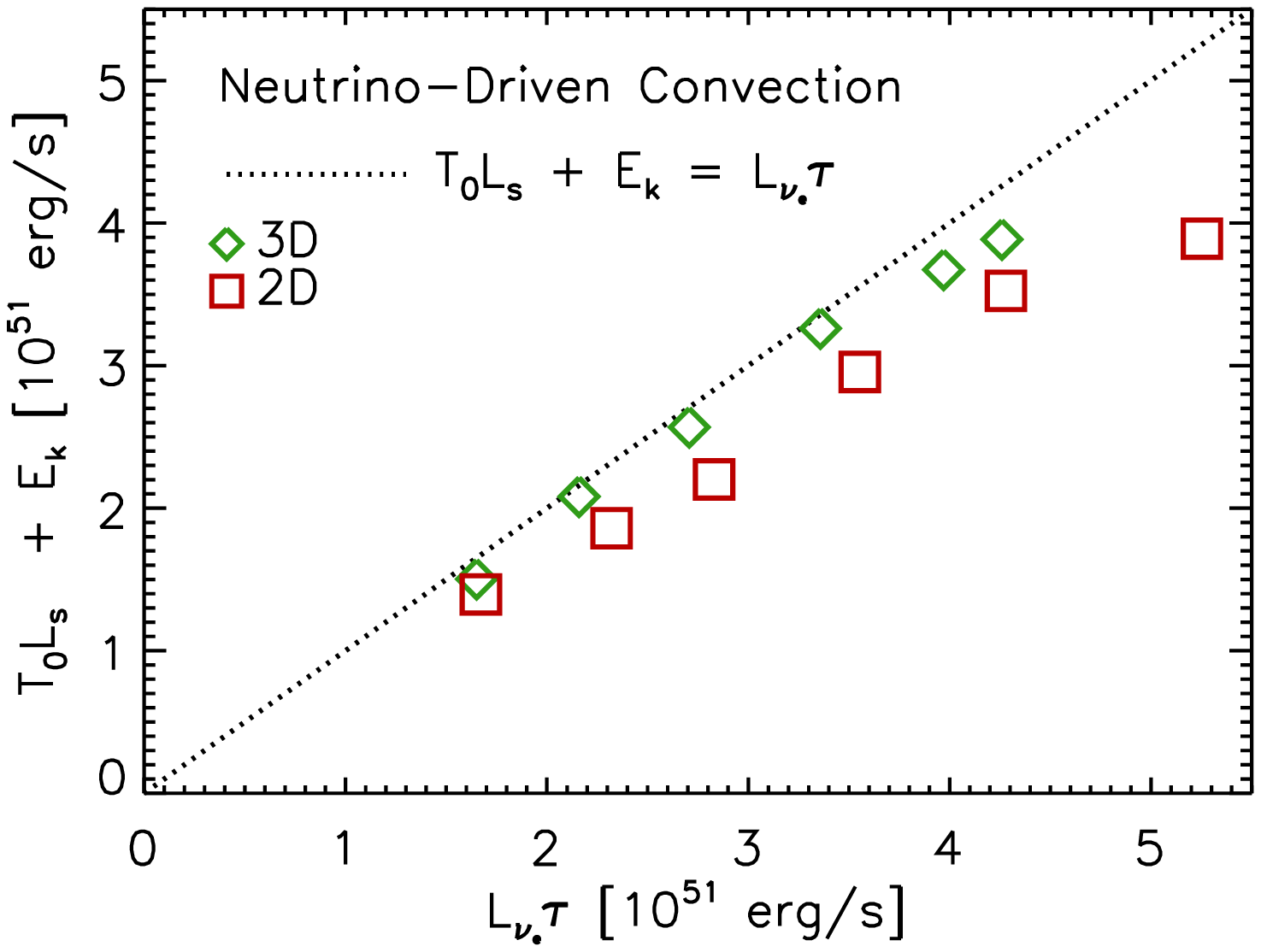}
\caption{Convective power ($T_0L_s + E_k$) vs. the driving neutrino
  power, $L_{\nu} \tau$.  Similar to Figure~\ref{tlsvslnutau}, the
  squares show the 2D simulation results, the
  diamonds show the 3D results, and the dashed line shows
  for 3D the expected linear scaling with driving neutrino power,
  Equation~(\ref{eq:tlsvslnutauek}). \label{tlsvslnutauek}}
\epsscale{1.0}
\end{figure*}

In Section~\ref{scalings}, we argue that the convective luminosity, $T_0
L_s$, and turbulent dissipation, $E_k$, are each proportional to
the driving neutrino-power, $L_{\nu} \tau$.  Furthermore, we propose
that the driving neutrino power is distributed between the convective
powers, i.e. $T_0 L_s + E_k \sim L_{\nu} \tau$.
Figures~\ref{tlsvslnutau}~and~\ref{tlsvslnutauek} compare these analytic
scalings, Equations~(\ref{eq:tlsvslnutau} and \ref{eq:tlsvslnutauek}), (dotted lines) with the results
from the 2D (squares) and 3D (diamonds) simulations.  Indeed, the turbulent luminosity
(Figure~\ref{tlsvslnutau}) is linearly proportional
to the driving neutrino power for both 2D and 3D.  Interestingly, the constant is lower for the 2D
simulations, suggesting that 2D is less efficient at driving a
convective entropy luminosity than 3D.  To calculate the constant of
proportionality for 3D, we merely report the ratio of the luminosities at
the last 3D point.  In Figure~\ref{tlsvslnutauek}, we find that the 2D
and 3D simulations confirm the hypothesis of Equation~(\ref{eq:tlsvslnutauek}) that
the neutrino-driving power is distributed among the convective
powers.  In summary, our simulations confirm our predictions that
turbulence scales with neutrino power, as is expected in buoyant
convection.  A result we could not predict, but that the simulations
tell us, is that convection is more vigorous for 3D than 2D for the
same luminosity.  In the Appendix, we show
that the scalings persist throughout the steady-state accretion phase.

\section{Expansion of Shock Radius due to Turbulence}
\label{shockexpansion}

An important aspect of the core-collapse problem is the presence of
the standing accretion shock, so we also consider how turbulence affects the stalled shock radius.  
Formally, the stalled shock is located where the upstream and the
downstream profiles satisfy the Rankin-Hugoniot jump conditions.  For
zero shock velocity, the mass flux, momentum flux, and energy flux
conditions are
\begin{equation}
\label{eq:rhjump1}
\Delta [\rho v] = 0 \, ,
\end{equation}
\begin{equation}
\label{eq:rhjump2}
\Delta [P + \rho v^2] = 0 \, ,
\end{equation}
and
\begin{equation}
\label{eq:rhjump3}
\Delta [\varepsilon + P/\rho + v^2/2] = 0\, .
\end{equation}
In detail, the shock position is a nontrivial solution to a boundary
value problem for $\rho$, $v$, and $P$.  However, with a few
reasonable approximations, the shock boundary condition can be reduced
to one expression.  First, we assume steady-state and that the mass accretion rate
($\dot{M}$) is constant.  Second, we assume that the upstream flow
is in near free-fall and essentially pressureless.  Third, we assume
for the purposes of this argument that the equation of state is
approximated by a $\gamma$-law, i.e. $P = (\gamma-1) \rho \varepsilon$.
The first and second assumptions completely determine
the upstream flow as a function of radius.  Because the upstream
flow is pressureless, we use the strong shock limit to determine the
shock compression ratio, i.e. $\rho_d/\rho_u
\approx (\gamma+1)/(\gamma-1)$.  Under these assumptions, the full
Rankine-Hugoniot jump conditions reduce to a single expression:
\begin{equation}
P_d = \rho_u v_u^2 \left ( 1 - \frac{\rho_u}{\rho_d}\right ) \, ,
\end{equation}
where $u$ ($d$) denotes upstream (downstream) state variables.
Since the term in parentheses is $\sim$1, the shock conditions reduce
to an expression that demands a balance between the upstream ram
pressure and the downstream thermal pressure.  
In essence, one can use the momentum jump condition
(Equation~\ref{eq:rhjump2}) to illuminate the
balancing conditions at the shock.
From here on, we focus on the momentum jump
condition with zero pressure on the upstream side, i.e.
\begin{equation}
\label{eq:jump1}
P_d + \rho_d v_d^2 \approx \rho_u v_u^2 \, .
\end{equation}

Using Reynolds decomposition, the momentum jump condition becomes
\begin{equation}
\label{eq:jump2}
P_d + \rho_d v_d^2 + \rho_d R_{rr} \approx \rho_u v_u^2 \, ,
\end{equation}
where the velocities, $v_d$ and $v_u$, are background velocities.
Hence, the new shock position is located where the post-shock thermal,
ram, and turbulent ram pressures balance the pre-shock ram pressure.
The addition of the turbulent ram pressure may result in larger
shock radii.

Equation~(\ref{eq:jump2}) by itself is not enough to determine the
shock position.  One must also specify the pre-shock and post-shock
profiles, and it is the intersection of these profiles that determines
the shock radius.  The pre-shock ram pressure is given by free-fall
assumptions, resulting in a fixed, relatively shallow profile
(e.g. $\rho_u v_u^2 \propto r^{-5/2}$).  The post-shock region is in
sonic contact and in rough hydrostatic equilibrium, so the postshock
pressure depends upon physics (such as cooling) of the entire
postshock region.  Fortunately, though, the post-shock
pressure profile can be expressed by a simple power-law (e.g. $P
\propto r^{-(3-4)}$), where the normalization depends upon the details
of cooling, etc.; we fit power-laws to the pre-shock and
post-shock profiles of 3D simulations and use Equation~(\ref{eq:jump2}) to
predict the average shock radius with and without turbulent ram
pressure.  We calculate the correct average shock radius only if we
include the turbulent ram pressure.

\begin{figure}[t]
\epssclone
\plotone{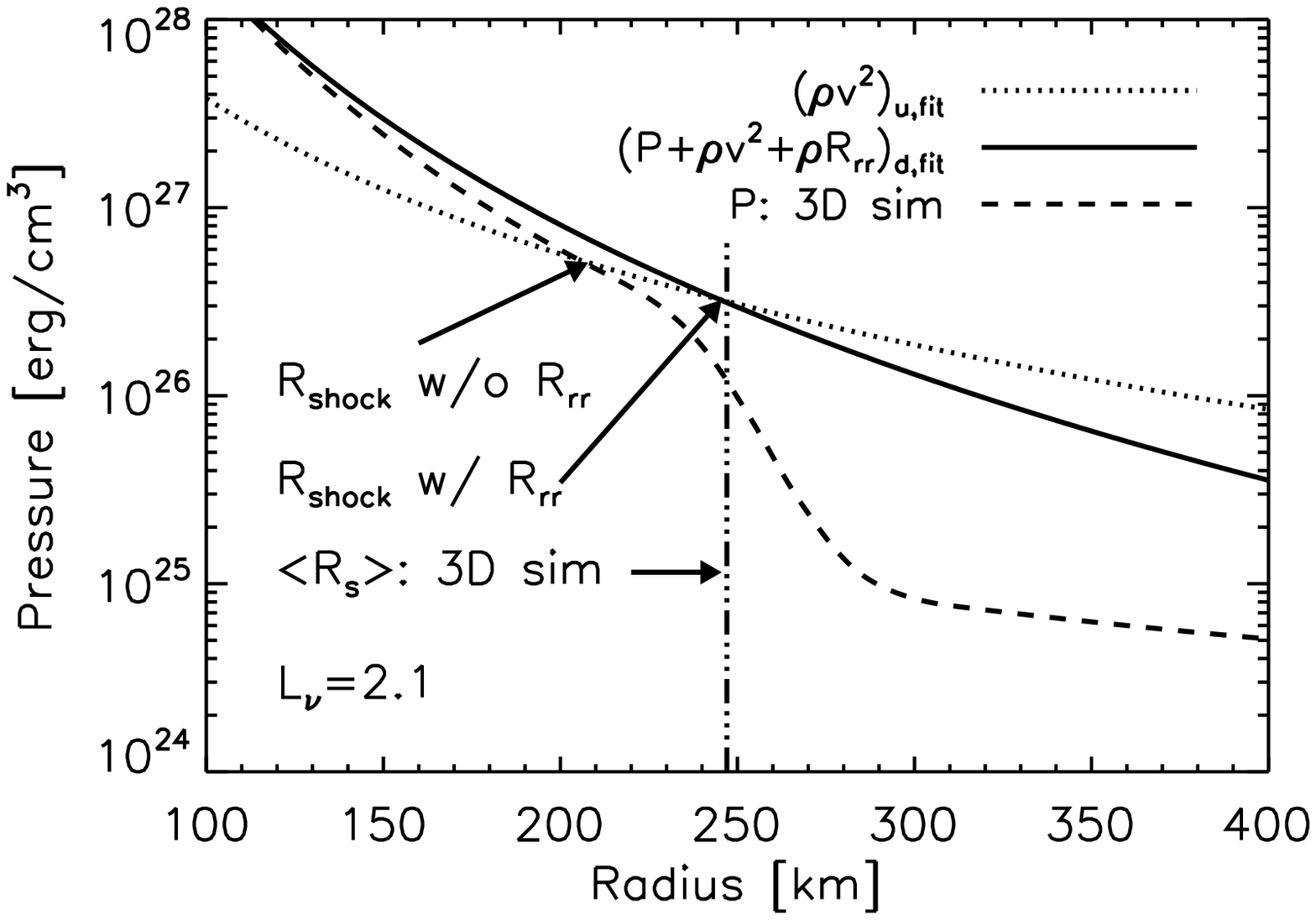}
\caption{Upstream (u) and downstream (d) momentum terms in the momentum shock condition, Equation~(\ref{eq:jump2}), and
  the resulting shock position at 250 ms after bounce.  In essence, the average shock radius is
  established where the ram pressure of the upstream flow (dotted line)
  balances the thermal, ram, and turbulent ram pressure of the
  downstream flow (solid line).  The solid and dotted lines are fits
  to their respective flows.  The dashed line shows the average
  pressure from the 3D simulations.  For comparison, we show the
  solid-angle-averaged shock radius ($\left < R_s \right >$) of the 3D simulation
  (dot-dot-dot-dashed line).  We calculate the shock radius with and
  without turbulent ram pressure and find that we can accurately
  estimate the shock radius only if we include the turbulent ram pressure.
  This demonstrates that the turbulent ram pressure is a
  non-negligible fraction of the total pressure.  
  In Figure \ref{rshockvslnue}, we use these curves to calculate the
  location of the average shock radius for several neutrino
  luminosities and show that it is a
  monotonically increasing function of $L_{\nu}$ and that turbulent
  pressure pushes the shock out to larger radii.  \label{shockcondition}}
\epsscale{1.0}
\end{figure}

\begin{figure*}[t]
\epssclone
\plotone{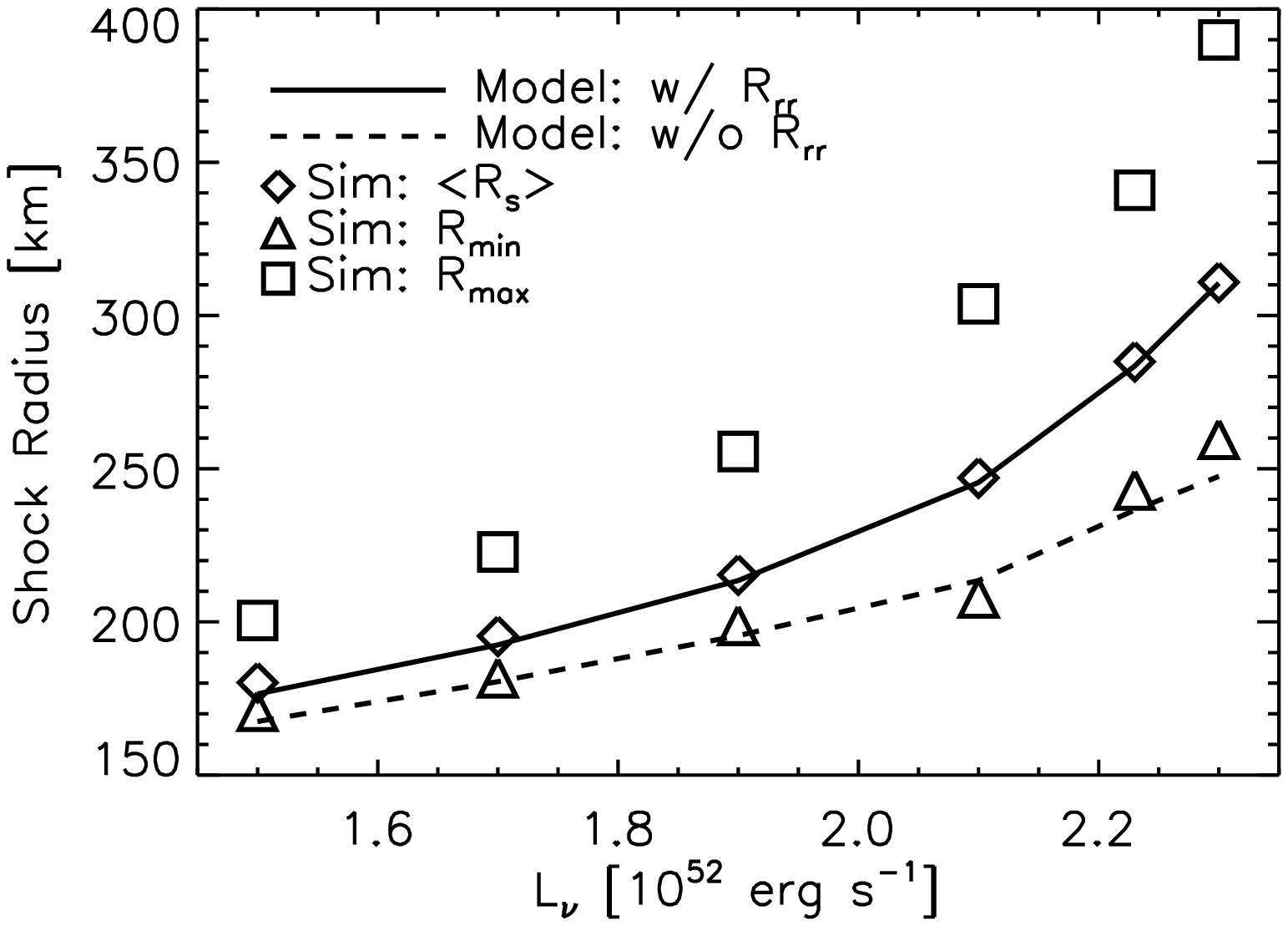}
\caption{Simulated and calculated shock radii, with and without turbulent
  ram pressure at 250 ms after bounce.  We plot the average (diamonds), minimum (triangles), and maximum
  (squares) shock radii vs. neutrino luminosity.
  Using the average pre- and post-shock thermal and momentum pressure
  profiles (Figure~\ref{shockcondition}), we calculate the expected shock radius
  with and without turbulent pressure.  Including the turbulent
  ram pressure gives a larger shock radius and matches the average
  shock radius from the simulations.  Calculations of the shock radii
  that exclude the turbulent ram pressure match the minimum shock radii of the
  3D simulations. \label{rshockvslnue}}
\epsscale{1.0}
\end{figure*}

Figures~\ref{shockcondition}~and~\ref{rshockvslnue} 
show that turbulent ram pressure explains in part the
expansion of the shock radius.  Figure
\ref{shockcondition} shows for one representative neutrino luminosity fits to the upstream ram pressure (dotted line) and
downstream thermal, ram, and turbulent pressures (solid line), as a function of
radius.  Using these pressure profiles, we calculate the shock to be
located where the upstream and downstream fits cross.  For comparison
we show the actual solid-angle-averaged shock radius
($\left < R_s \right > =  \int R_s(\theta,\phi) \, d \Omega/( 4 \pi ) $) from a 3D simulation (dot-dot-dot-dashed
line).  Including the turbulent pressure leads to a more accurate
prediction of the shock radius.  If we neglect the turbulent pressure,
then we estimate a shock radius that is smaller by about 40 km
(labeled ``$R_{\rm shock}$ w/o $R_{rr}$'').  To be clear, we are
not concluding that this is where the shock would be located in the
absence of turbulent pressure (i.e. a 1D simulation, although it is
quite close).  We are merely demonstrating that the turbulent ram
pressure is a sizable fraction of the microscopic pressure and that
one can not ignore the importance of the turbulent pressure.

Figure \ref{rshockvslnue} shows the resulting shock locations as a
function of driving neutrino luminosity.  This plot shows the modeled
shock radii, including turbulent ram pressure (solid line) and the
modeled shock radii excluding turbulent ram pressure (dashed line).  For comparison, we show the calculated minimum (triangles),
average (diamonds), and maximum (squares) shock radii for 3D
simulations, all at 250 ms after bounce\footnote{In the Appendix, we
  plot these results at other times after bounce and find that our results and
  conclusions hold throughout the steady state accretion phase.}.  Including turbulent
pressure in the post-shock profile predicts shock radii that agree
with the measured average shock radius.  Excluding the
turbulent pressure under-predicts the average shock radius.
On the other hand, excluding the turbulent pressure gives shock radius
predictions that are consistent with the minimum shock radii.  This
suggests that the minimum shock radii occur at places and times where
the fluctuating turbulent motions are instantaneously negligible.  On
average though, the turbulent motions are not negligible and influence
the average shock radius.  

Showing that turbulent ram pressure causes expansion of the shock radius
does not by itself prove that buoyant-driven convection is responsible for the
expansion.  Any instability that leads to turbulence would give a
similar prediction.  However, the dependency of the shock radii and the
turbulent pressure on $L_{\nu}$ does strongly suggest the prominence of
neutrino-driven convection.

\section{Conclusions and Discussion}
\label{conclusions}

We have identified in Section~\ref{scalings} four ways in which
the aspherical, nonlinear flow of 3D CCSN simulations is consistent
with buoyancy-driven convection.  First, the turbulent
luminosity is positive in the gain region where buoyancy actively
drives convection, and the turbulent luminosity is negative in the stably-stratified region where
buoyancy decelerates convective plumes.  Second, the radial
component of the Reynolds stress is in rough equipartition with the
tangential components, i.e. $R_{rr} \sim R_{\theta \theta} + R_{\phi
  \phi}$; this result is observed in other contexts of buoyancy-driven
convection and is expected when convection is driven radially, but
dissipated among all the components \citep{arnett09,garaud10,murphy11b}.  Third, we find that
turbulent dissipation is balanced by buoyant driving.
Fourth, both the turbulent luminosity and turbulent dissipation
scale with the driving neutrino power.
In essence, these results are consistent with the hypothesis that
neutrino-driven convection is the dominant multi-dimensional
instability in our 3D CCSN simulations.

Since these results are consistent with the buoyancy-driven
hypothesis, one wonders if the SASI is suppressed during
neutrino-driven explosions.
Of course, our results
do not prove that the SASI is absent.  Rather, they suggest that the
nonlinear motions are merely consistent with buoyant convection and
that if the SASI is present it mimics buoyant convection, or is subdominant.  Numerous analytic and numerical studies
have shown that if the conditions are right, a nonlinear SASI
arises.  However, these studies were performed largely in the
absence of neutrino heating or used high accretion rates and small
initial perturbations.  A few did include neutrinos, and given
sufficient initial perturbations, even in
these investigations neutrino-driven convection seems eventually to
dominate \citep{foglizzo06,scheck08,fernandez09,mueller12,hanke13}.  It appears that in attempting to
isolate the SASI mechanism, researchers were suppressing the dominant
nonlinear instability in core-collapse simulations.

Minimally, there are two hypotheses for the dominant multi-dimensional
instability: the SASI and buoyancy-driven convection.
Our results suggest that when neutrinos drive explosions, they also
drive convection and turbulence.  However, there is as yet no self-consistent theory for a
nonlinear SASI mechanism, so we cannot test the SASI at this time.
In fact, it is entirely possible that convective and SASI theories
might predict similar characteristics.  For example, in both cases, turbulence is most
likely dissipated in accord with Kolmogorov's hypothesis.  However, we have shown
that turbulent dissipation is balanced by buoyant driving, which seems
an unlikely prediction of a nonlinear SASI theory.  In any case, it is
clear that a nonlinear theory for the SASI must be developed before
we can definitively claim that the SASI is subdominant.  Furthermore,
even though our approximations are designed to closely mimic more
self-consistent simulations, a robust conclusion on the importance of
convection must wait for full 3D neutrino-transport hydrodynamic
simulations.  

In the meantime, we have shown that the turbulence in our 3D
simulations is consistent with buoyant convection.  Not
surprisingly, we find that neutrino-driven convection accompanies
neutrino-driven explosions.

\section*{Acknowledgments}
The authors acknowledge stimulating interactions with Jason Nordhaus,
Ann Almgren, and Thomas Janka. A.B. is supported by the Scientific
Discovery through Advanced Computing (SciDAC) program of the DOE,
under grant number DE-FG02-08ER41544, the NSF under the subaward
no. ND201387 to the Joint Institute for Nuclear Astrophysics (JINA,
NSF PHY-0822648), and the NSF PetaApps program, under award
OCI-0905046 via a subaward no. 44592 from Louisiana State University
to Princeton University. The authors would like to thank the members
of the Center for Computational Sciences and Engineering (CCSE) at
LBNL for their invaluable support for CASTRO. The authors employed
computational resources provided by the TIGRESS high performance
computer center at Princeton University, which is jointly supported by
the Princeton Institute for Computational Science and Engineering
(PICSciE) and the Princeton University Office of Information
Technology; by the National Energy Research Scientific Computing
Center (NERSC), which is supported by the Office of Science of the US
Department of Energy under contract DE-AC03-76SF00098; and on the
Kraken supercomputer, hosted at NICS and provided by the National
Science Foundation through the TeraGrid Advanced Support Program under
grant number TG-AST100001.

\begin{appendix}
\label{appendix}

In this paper, we address which mechanism drives turbulence during the
stalled accretion phase and thereby sets the stage for explosion.  In
this appendix, we consider the validity of our conclusions as a
function of time.  We find that the temporal behavior falls into two
distinct classes: those that explode within 1 s after bounce and
those that don not.  Rather than showing the time evolution for all
neutrino luminosities, we present only two neutrino luminosities; one
represents a non-exploding model ($L_{\nu}= 2.1$), and one represents
an exploding model ($L_{\nu}=2.23$).
As is expected, we find that our conclusions are valid during the steady-state
accretion phases and are less valid during the dynamic phases, which
include the initial phase when the shock is settling and the
explosive phase.

\begin{figure*}[t]
\epsscale{0.7}
\plotone{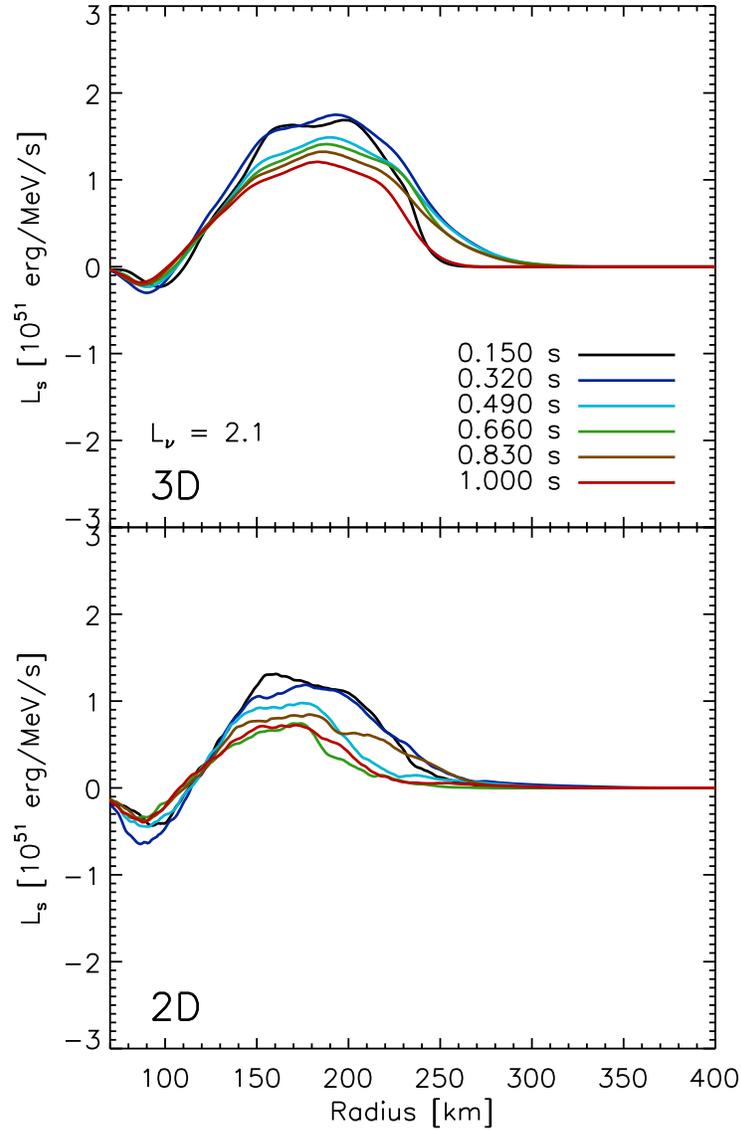}
\caption{Turbulent entropy luminosity , $L_s = 4 \pi r^2 \left < F_s
  \right >$ vs. radius and time after bounce for 3D (top panel) and
  2D (bottom panel) simulations.  $L_{\nu}=2.1$ represents a
  non-exploding model. This figure shows that 3D and 2D simulations
  are consistent with a neutrino-driven convection hypothesis through
  out the steady-state accretion phase.  In particular, during this
  phase, the entropy
  luminosity remains positive where buoyancy actively drives
  convection and negative where the stable layers decellerate the
  convective plumes.\label{lsprofilesvstime_2.1}}
\epsscale{1.0}
\end{figure*}

\begin{figure*}[t]
\epsscale{0.7}
\plotone{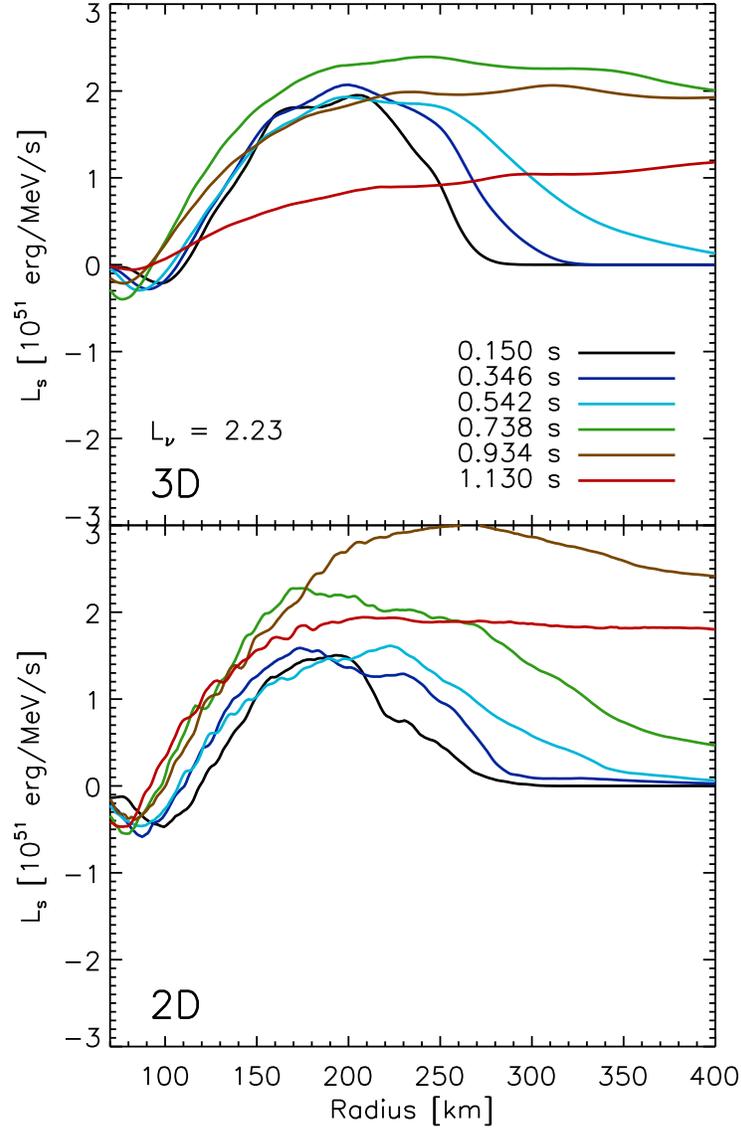}
\caption{Similar to Figure~\ref{lsprofilesvstime_2.1} except for an
  exploding model ($L_{\nu} = 2.23$).  Once again, during the
  steady-state accretion phase, the convective entropy luminosity is
  positive where convection is driven by buoyancy and negative where
  it is decellerated by buoyancy.  For the times during explosion
  ($t > 0.6$ s), the entropy luminosity first rises and then falls. \label{lsprofilesvstime_2.23}}
\epsscale{1.0}
\end{figure*}

\begin{figure*}[t]
\epsscale{0.7}
\plotone{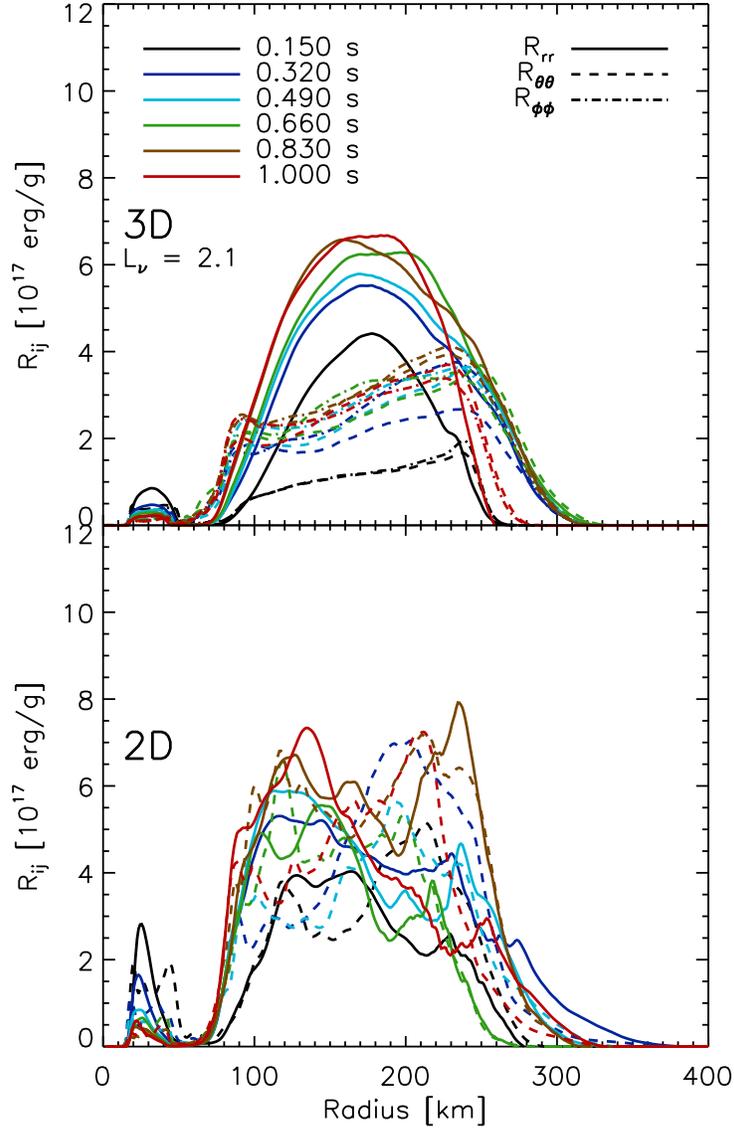}
\caption{Reynolds stress as a function of radius radius and time after bounce for 3D (top panel) and
  2D (bottom panel) simulations.  As in Figure~\ref{lsprofilesvstime_2.1}, we
  choose to show $L_{\nu}=2.1$ to represent a
  non-exploding model. 
  For 3D, $R_{rr} \sim R_{\theta \theta} +
  R_{\phi \phi}$ globally, and $R_{\phi \phi} \approx R_{\theta
    \theta}$ locally for all times.  For 2D, $R_{rr}$ is in rough
  equipartition with the only tangential compoenent, $R_{\theta
    \theta}$.  Hence, both 3D and 2D are consistent with the
  expectations of buoyantly driven convection at all times. \label{reyprofilesvstime_2.1}}
\epsscale{1.0}
\end{figure*}

\begin{figure*}[t]
\epsscale{0.7}
\plotone{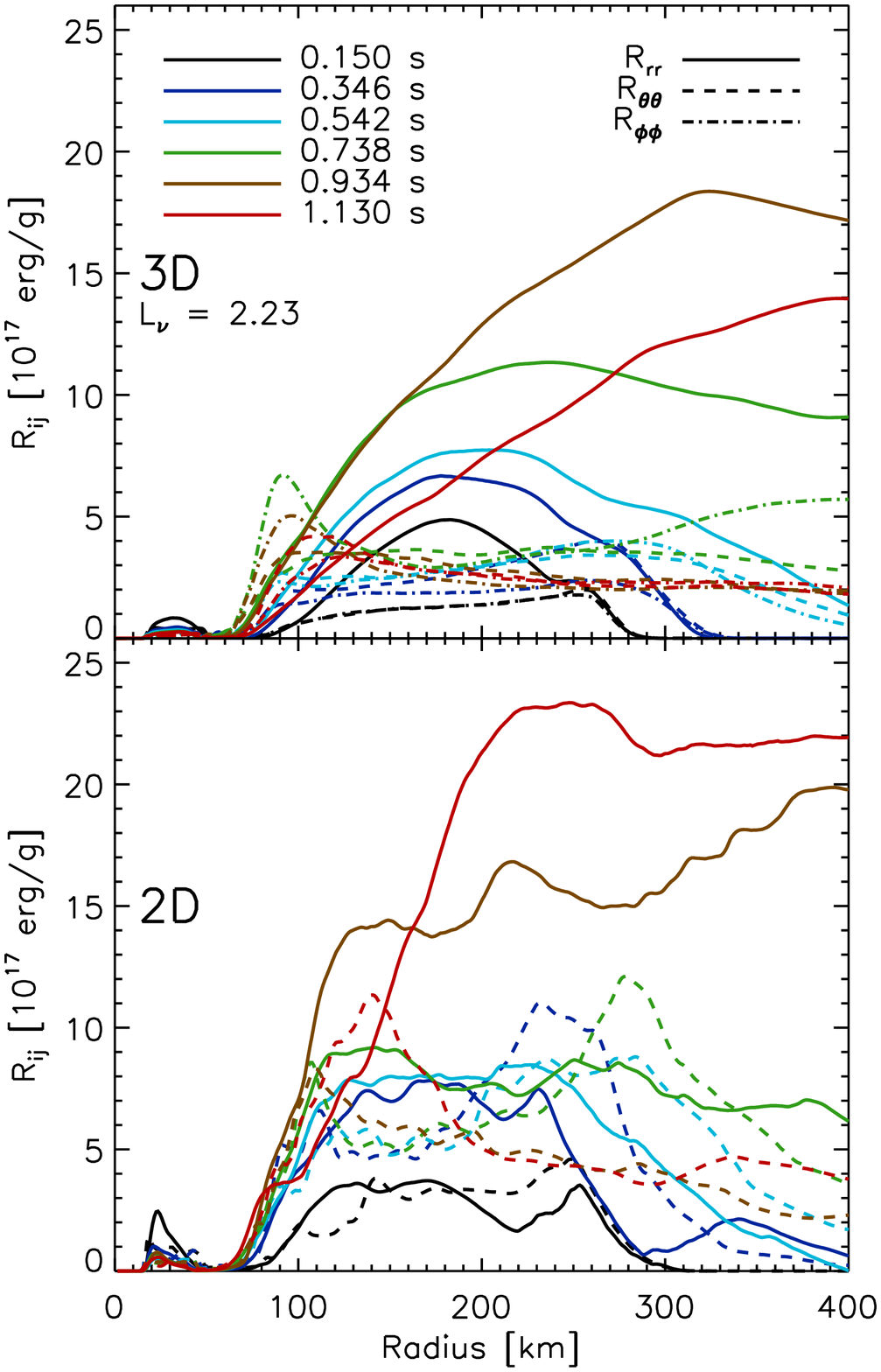}
\caption{Similar to Figure~\ref{reyprofilesvstime_2.1} except for an
  exploding model ($L_{\nu} = 2.23$).  Before explosion, the
  steady-state profiles are consistent with buoyant driven
  convection.  During explosion, the radial component grows significantly.\label{reyprofilesvstime_2.23}}
\epsscale{1.0}
\end{figure*}

Figures \ref{lsprofilesvstime_2.1}-\ref{reyprofilesvstime_2.23} show
the temporal evolution of the turbulent entropy luminosity and the
Reynolds' stress for the 2D and 3D simulations.  During the
steady-state accretion phase, both measures of turbulence are
consistent with buoyant driven convection.  The turbulent entropy
luminosity is positive where buoyancy actively drives turbulence and
negative where the stable layer decelerates the convective plumes.
Furthermore, the distribution of Reynolds' stresses is roughly in
equipartition between the radial and tangential components and is a
hallmark of buoyancy driving in the radial direction. For both the
non-exploding and exploding model, the convective velocities as
measured by the Reynolds' stress increases with time, especially
during the steady-state accretion phase.  However, the convective
entropy luminosity decreases with time for the non-exploding model and
is non-monotonic for the exploding model.

\begin{figure*}[t]
\epssclone
\plotone{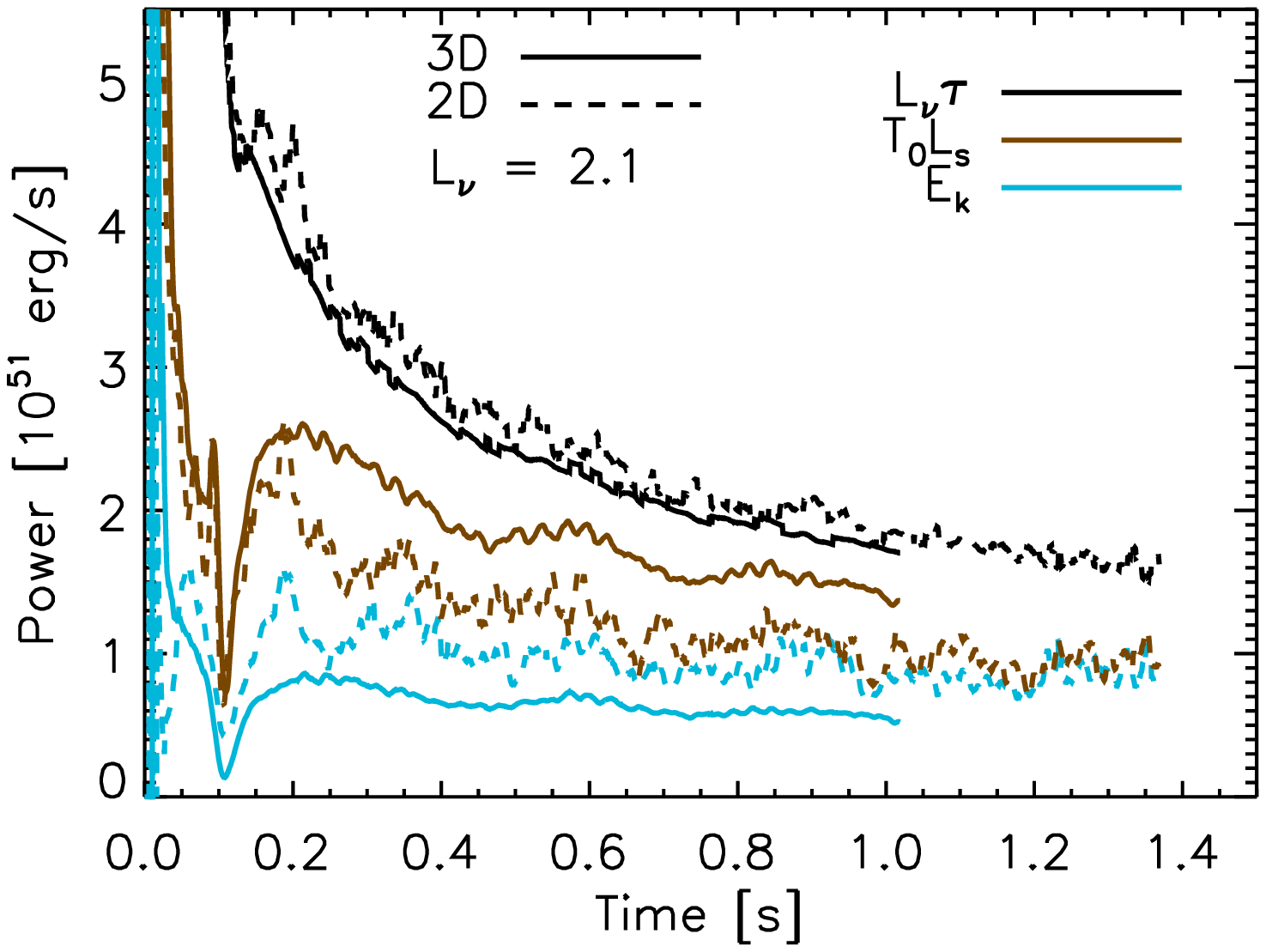}
\caption{Driving neutrino power ($L_{\nu} \tau$) and convective power
  ($T_0L_s$ and $E_k$) vs. time after bounce for representative
  2D (dashed lines) and 3D (solid lines) non-exploding models
  ($L_{\nu} = 2.1$).  2D turbulent dissipation ($E_k$) is higher than 3D
  turbulent dissipation, which is due to higher radial velocities in
  2D.  Conversely, the turbulent convective luminosity is higher in
  3D than in 2D.  This is likely a result of longer dwell times in 3D \citep{dolence13}.  \label{integralsvstime_2.1}}
\epsscale{1.0}
\end{figure*}

\begin{figure*}[t]
\epssclone
\plotone{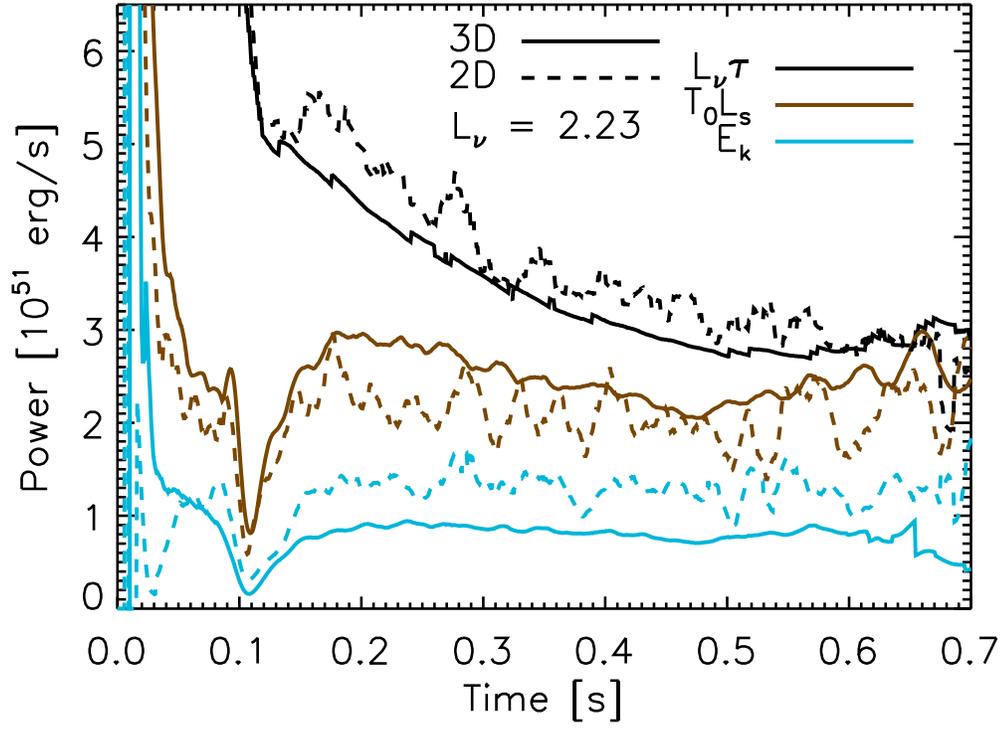}
\caption{Similar to Figure~\ref{integralsvstime_2.1} except for
  representative exploding models ($L_{\nu} = 2.23$).  During the
  steady-state accretion phase ($0.15 < t < 0.6$) the orderings and
  evolution of the power is similar to those in the non-exploding
  models (Figure~\ref{integralsvstime_2.1}).   \label{integralsvstime_2.23}}
\epsscale{1.0}
\end{figure*}

\begin{figure*}[t]
\epssclone
\plotone{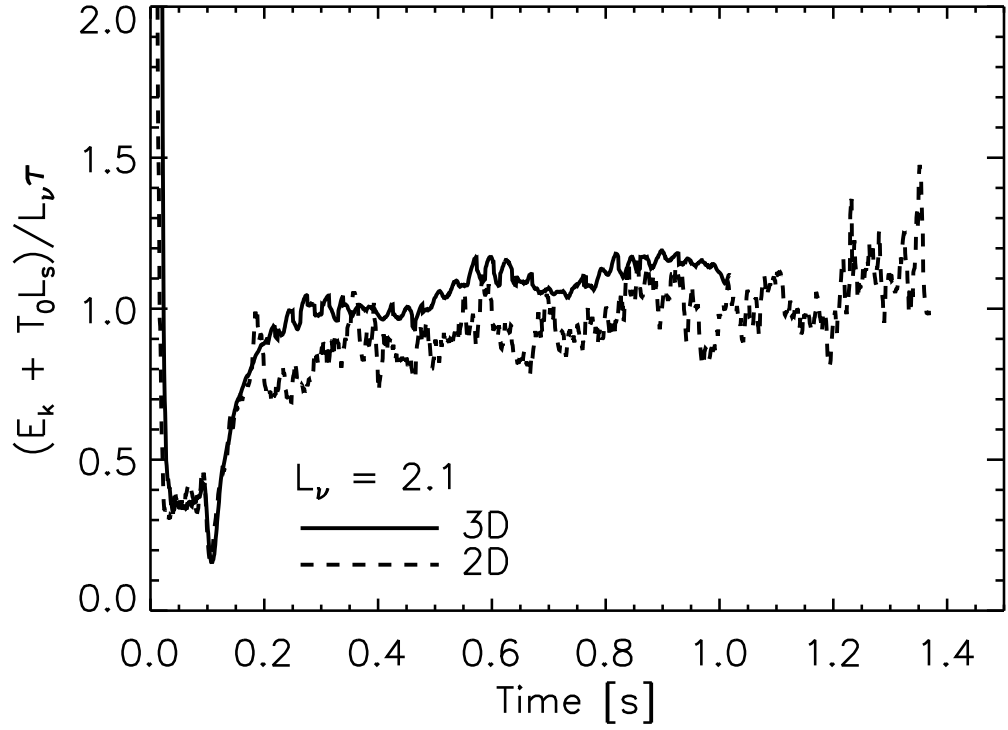}
\caption{Power ratio vs. time after bounce for the representative
  non-exploding models ($L_{\nu} = 2.1$).  Despite the fact that the
  driving neutrino power ($L_{\nu} \tau$) varies by a factor of four
  during the steady-state accretion phase
  (Figure~\ref{integralsvstime_2.1}), the ratio of the convective
  power to the neutrino power is of order one during this entire
  phase.  The fact that this scaling persists both over a range of
  luminosities (Figure~\ref{tlsvslnutauek}) and over time is a strong
  indicator that the turbulent motions are buoyantly driven by
  neutrinos. \label{powerratiovstime_2.1}}
\epsscale{1.0}
\end{figure*}

\begin{figure*}[t]
\epssclone
\plotone{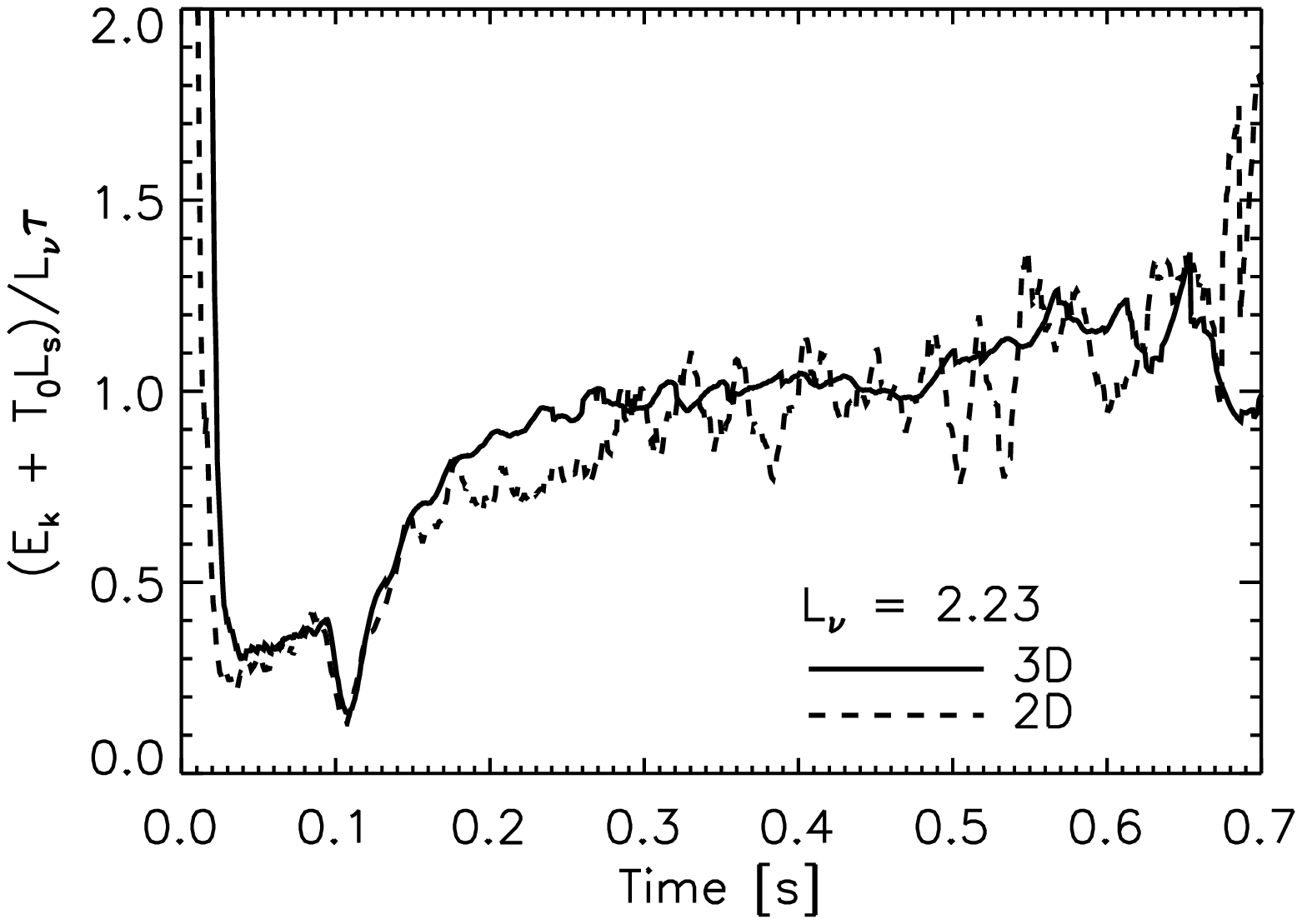}
\caption{Power ratio vs. time after bounce for the representative
  exploding models ($L_{\nu} = 2.23$).  As in the case for the
non-exploding models (Figure~\ref{powerratiovstime_2.1}), the ratio
of the convective power to the neutrino power is of order one during
the steady-state accretion phase, and once again, the fact that this scaling persists both over a range of
  luminosities (Figure~\ref{tlsvslnutauek}) and over time is a strong
  indicator that the turbulent motions are buoyantly driven by
  neutrinos.   \label{powerratiovstime_2.23}}
\epsscale{1.0}
\end{figure*}

In Figures \ref{integralsvstime_2.1}-\ref{powerratiovstime_2.23}, we
demonstrate that the convetive power scales with the driving neutrino power, $L_{\nu}\tau$, throughout the steady-state accretion
phase.  First, we plot in Figures \ref{integralsvstime_2.1} \& \ref{integralsvstime_2.23} the driving neutrino power ($L_{\nu}\tau$), the
convective luminosity ($T_0L_s$), and the turbulent dissipation rate
($E_k$) versus time after bounce.  We show both 2D and 3D results for
the non-exploding ($L_{\nu}\tau = 2.1$, Figure
\ref{integralsvstime_2.1}) and exploding ($L_{\nu}\tau = 2.23$,
Figure~\ref{integralsvstime_2.23}) models.  Comparing these measures
of power is most meaningful during the steady-state accretion phase:
later than $\sim$0.15 s and earlier than explosion.  The $L_{\nu}$
does not explode during the calculation, so we consider the full run.
The $L_{\nu} = 2.23$ model, on the other hand, explodes at $\sim$0.6 s after bounce.
Therefore, we do not plot the powers beyond 0.65 s.  Beyond this time,
these powers, especially $E_k$, are ill defined, difficult to
calculate, and confusing to interpret.  Finally,
in Figures~\ref{powerratiovstime_2.1}~and~\ref{powerratiovstime_2.23},
we plot the ratio of convective power to driving neutrino power, which
we expect to be of order one.

In general, Figures~\ref{integralsvstime_2.1}-\ref{powerratiovstime_2.23} show that the scaling relations
presented in this paper persist during the steady accretion rate
phase.
Figures~\ref{integralsvstime_2.1}~and~\ref{integralsvstime_2.23} show
that the turbulent dissipation is higher for 2D than
3D, but that the turbulent entropy luminosity is higher for 3D than
2D.
The first result is a consequence of the fact that the radial
convective velocities are higher in 2D than 3D.  Given the lower
convective velocities in 3D, the material can dwell longer in the gain
region; this is possibly the explanation for the larger convective
entropy luminosity.  The most striking result of
Figures~\ref{integralsvstime_2.1}-\ref{powerratiovstime_2.23} is that
despite the fact that the driving neutrino power varies by a large
factor (4 for $L_{\nu} = 2.1$), the ratio remains of order one during
the entire steady-state accretion phase.  In other words, the scaling
relation between the driving power and the convective powers
persists.

\begin{figure*}[t]
\epssclone
\plotone{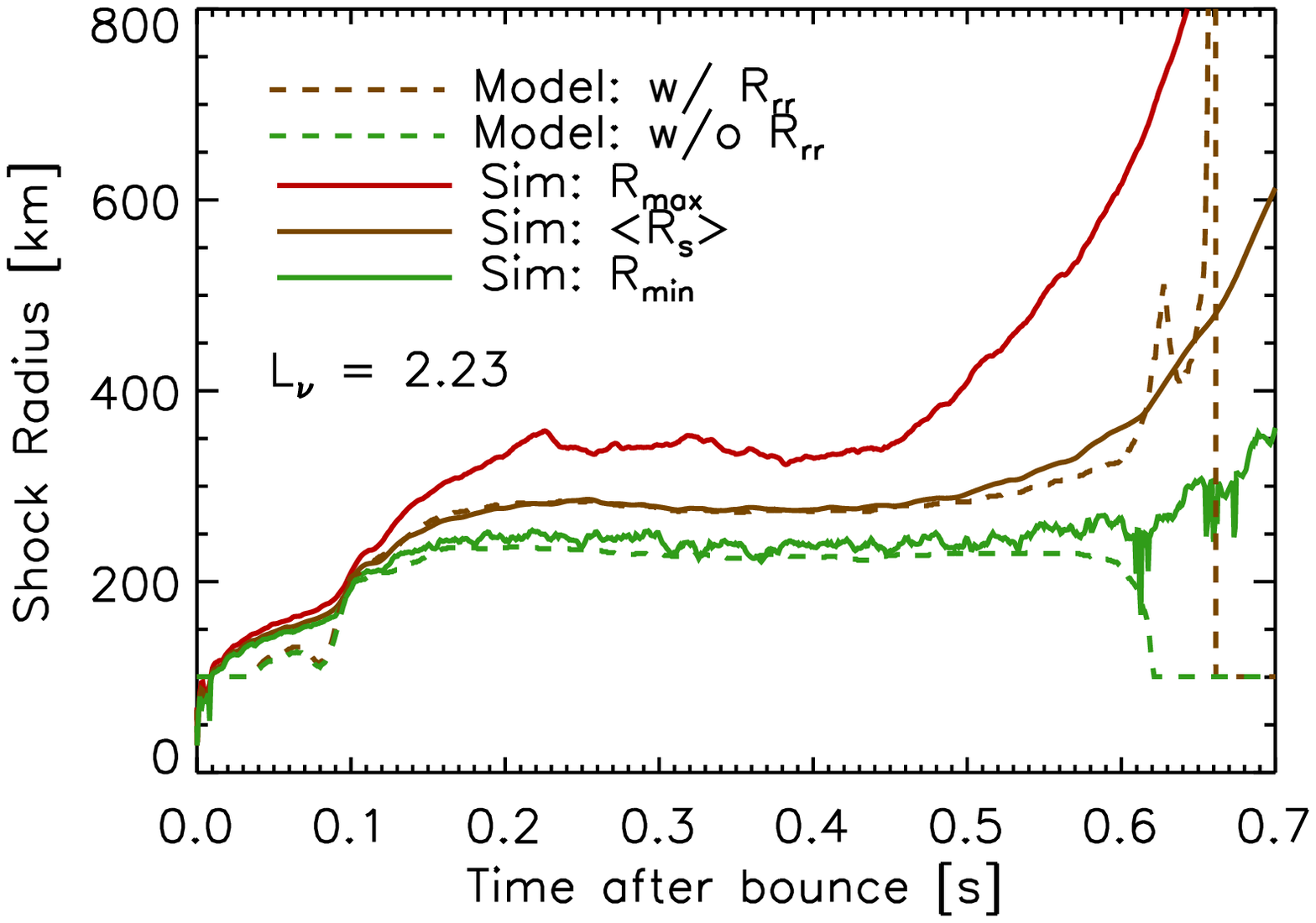}
\caption{Simulated and calculated shock radii, with and without turbulent
  ram pressure as a function of time after bounce for the $L_{\nu}=
  2.23$ model.  Here we find that the calculated shock radii
  accurately represent the simulations during the steady-state phases.
  Before the shock settles ($t < 0.1$s) and after explosion begins
  ($t> 0.6$s), the steady-state model is naturally inaccurate.  During
  the bulk of the steady-state phase, though, the calculated shock
  radii are accurate, and confirm that turbulent pressure is
  important. \label{rshockvslnue_vstime}}
\epsscale{1.0}
\end{figure*}

Finally, Figure~\ref{rshockvslnue_vstime} shows that our model for
calculating the shock radii (with and without turbulent pressure) is
valid for the entire steady-state accretion phase.

\end{appendix}


\end{document}